\documentclass[hyper,12pt]{JHEP3} % 10pt is ignored!

\usepackage{epsfig}
\usepackage{amsmath}
\usepackage{cite}

\newcommand{\beq}{\begin{equation}}
\newcommand{\eeq}{\end{equation}}

\title{Connecting LHC, ILC, and Quintessence}

\author{Daniel J.~H.~Chung, Lisa L.~Everett\\
        Department of Physics, University of Wisconsin, 
        Madison, WI 53706, USA\\
        E-mail: \email{danielchung@wisc.edu, leverett@wisc.edu}}
\author{Kyoungchul Kong\\
        Theoretical Physics Department, Fermilab,
        Batavia, IL 60510, USA\\
        E-mail: \email{kckong@fnal.gov}}
\author{Konstantin T.~Matchev\\
        Institute for Fundamental Theory, University of Florida,
        Gainesville, FL 32611, USA\\
        E-mail: \email{matchev@phys.ufl.edu}}

\received{\today}               %%
\accepted{\today}               %% These are for published papers.

\preprint{MADPH-07-1493 \\
          FERMILAB-PUB-07-194-T\\
          UFIFT-HEP-07-7 \\
          June 13, 2007
          } % OR: \preprint{Aaaa/Mm/Yy\\Aaa-aa/Nnnnnn}
\abstract{If the cold dark matter consists of weakly interacting
massive particles (WIMPs), anticipated measurements of the WIMP
properties at the Large Hadron Collider (LHC) and the International
Linear Collider (ILC) will provide an unprecedented experimental probe
of cosmology at temperatures of order 1 GeV.  It is worth emphasizing
that the expected outcome of these tests may or may not be consistent
with the picture of standard cosmology.  For example, in
kination-dominated quintessence models of dark energy, the dark matter
relic abundance can be significantly enhanced compared to that
obtained from freeze out in a radiation-dominated universe. Collider
measurements then will simultaneously probe both dark matter and dark
energy. In this article, we investigate the precision to which the LHC
and ILC can determine the dark matter and dark energy parameters under
those circumstances.  We use an illustrative set of four benchmark
points in minimal supergravity in analogy with the four LCC benchmark
points. The precision achievable together at the LHC and ILC is
sufficient to discover kination-dominated quintessence, under the
assumption that the WIMPs are the only dark matter component.  The LHC
and ILC can thus play important roles as alternative probes of both
dark matter and dark energy.  }


\begin{document} 

%%%%%%%%Section #1  %%%%%%%%%%%%%%%%%%%%%%%%%%%%%%%%%%%
%%%%%%%%%%%%%%%%%%%%%%%%%%%%%%%%%%%%%%%%%%%%%%%%%%%%%%%
\section{Introduction}
\label{sec:intro}

Current cosmological data
\cite{Spergel:2006hy,Astier:2005qq,Eisenstein:2005su,Riess:2004nr}
indicate that the energy density of the universe today is dominated by
degrees of freedom beyond those of the Standard Model (SM) of particle
physics, with approximately 73\% given by dark energy, a form of
energy characterized by negative pressure, and approximately 23\%
given by nonbaryonic dark matter.  Given that future colliders such as
the Large Hadron Collider (LHC) and International Linear Collider
(ILC) are designed to probe physics beyond the SM, it is important to
investigate whether and to what extent these experiments can probe
these elusive forms of matter and energy.

For dark matter, it is well known that both direct and indirect collider
physics connections can be established, depending on the properties of
the dark matter candidate.  One particularly well-motivated class of
models is that in which the dark matter is a neutral weakly interacting
massive particle (WIMP). Such
particles are in thermal equilibrium in the early universe, and the
standard freeze out calculation predicts a value for their relic density
which is in the right ballpark as required by cosmology.  WIMPs are
naturally present in models which attempt to solve the gauge hierarchy
problem, since such models generically introduce new weakly
interacting particles.  Prototype scenarios include
models with softly broken $N=1$ supersymmetry (SUSY), in which there is
a conserved discrete symmetry, R-parity. In such models the lightest
superpartner (LSP), typically a neutralino, is a viable cold dark matter
candidate.  The connections between astroparticle and collider physics
have been extensively explored in the literature on supersymmetric dark
matter 
\cite{Polesello:2004qy,Battaglia:2004mp,Allanach:2004xn,Battaglia:2004gk,
deAustri:2006pe,Bourjaily:2005ax,Belanger:2005jk,Moroi:2005nc,
Birkedal:2005jq,Battaglia:2005ie,Moroi:2005zx,
Birkedal:2005aa,
Nojiri:2005ph,Baltz:2006fm,Drees:2007hr,
Djouadi:2006pg,White:2006wh,Kraml:2006au}.
More recently, other WIMP candidates have emerged in new physics
scenarios which include a conserved discrete symmetry, such as models
with flat \cite{Servant:2002aq,Cheng:2002ej} or warped
\cite{Agashe:2004ci,Agashe:2004bm} extra dimensions, and Little Higgs
models
\cite{Birkedal-Hansen:2003mp,Cheng:2003ju,Katz:2003sn,
Cheng:2004yc,Birkedal:2006fz}.

It is much more difficult to establish a direct link to collider
physics for the case of dynamical dark energy, since it is typically
far more weakly coupled to the SM (for an example of an intriguing
exception, see \cite{Stojkovic:2007dw}).  Although this energy density
can be the cosmological constant, in which case cosmology has given us
an invaluable clue to the cosmological constant problem, it is
plausible that the dark energy can be associated with an effective
scalar field degree of freedom, commonly called a quintessence field
\cite{Caldwell:1997ii,Wetterich:1987fk,Peebles:1987ek,Freese:1986dd,bertolami}.
Collider experiments are unlikely to probe the unknown quintessence
field directly, given its typical range of masses and couplings.
Connections between collider physics and dark energy in the form of
quintessence are thus necessarily indirect, and only a few are known.
One connection is in the context of models with low energy
supersymmetry.  Since four-dimensional de Sitter space does not admit
unbroken $N=1$ supersymmetry \cite{Nahm:1977tg}, the discovery of
supersymmetry through measurements of spins and couplings, and hints
for its spontaneous breaking mechanism through measurements of mass
patterns, will give us clues to the cosmological constant problem and
possibly its associated quintessence dynamics.  Other indirect
connections with collider physics result because quintessence dynamics
can leave an imprint on the cosmological history, which in turn
determines an observable quantity that depends partly on parameters
deduced from collider measurements.

One intriguing collider physics connection can arise through the
gravitational interactions of the dark matter with quintessence.  For
thermal relics such as LSP dark matter, freeze out occurs when the
expansion rate becomes larger than the interaction rate for reactions
which change the LSP number density.  In the standard cosmological
scenario, freeze out represents the gravitational interactions between
the cold dark matter and the relativistic gas of particles dominating
the energy density of the universe at the time of freeze out.  However, if the
coherent field energy density of the quintessence field dominates during freeze out, freeze out now represents the gravitational interactions between the dark matter and
the dark energy field degrees of freedom.  In such cases, the relic
density of the WIMP can be very strongly affected.  More explicitly, the
WIMP energy density today can be written as
\begin{equation}
\Omega_\chi h^2 \propto \left( \frac{T_{\rm today}}{m_\chi
  x_F} \right)^3 \left( \frac{m_\chi H_F}{\langle \sigma_A v\rangle}
  \right),
\label{danrelation}
\end{equation}
where $H_F$ is the Hubble expansion rate at freeze out, $m_\chi$ is the
mass of the WIMP, $\langle \sigma_A v\rangle$ is the
thermally-averaged annihilation cross-section, $x_F \equiv T/m_\chi
\sim 1/20$ for electroweak scale cross-sections (with only logarithmic
dependences on $H_F$, $\langle \sigma_A v\rangle$, and $m_\chi$), and
$T_{\rm today}$ is the temperature of the cosmic microwave
background (CMB) today.  Eq.~(\ref{danrelation}) demonstrates that if
$H_F$ is increased due to the quintessence energy density, the WIMP relic
density can be enhanced for the same microphysics contained in
$\langle \sigma_A v\rangle$.  Note that Eq.~(\ref{danrelation}) also
makes manifest the well-known property in standard cosmology that
$\Omega_\chi h^2$ becomes approximately independent of the mass
$m_\chi$ for a fixed $\langle \sigma_A v\rangle$, since in this case
$H_F \sim x_F^2 m_\chi^2/M_{pl}$.

A challenge for such alternate scenarios is that standard big bang
nucleosynthesis (BBN) constrains any extra contribution to the
relativistic energy density. The quintessence energy density therefore
has to be large during the time of WIMP freeze out ($T \sim O(1)$ GeV) and
then dilute more quickly than radiation to become effectively
irrelevant by the time that BBN commences ($T_0\sim 10^{-3}$ GeV).  As
pointed out by Salati \cite{Salati:2002md}, this can occur for
scenarios in which the universe is driven by the quintessence kinetic
energy (the kination-dominated period) before BBN.  (Related
scenarios were also suggested before by
\cite{Kamionkowski:1990ni,Barrow:1982ei}.)  Long
after BBN, the universe can enter a
quintessence potential energy dominated regime.  Such kination 
dominated freeze out scenarios are then consistent with standard 
cosmology and predict that the {\em standard}
relic abundance computed from the parameters extracted from the next
generation of colliders will be mismatched from the relic abundance
deduced from observational cosmology.  Indeed, the kination hypothesis
is only one of many possibilities we must turn to if future
experiments find a mismatch between the standard computations of 
the relic abundance and cosmological observations.

The implications of kination domination for LSP dark matter have been
explored previously \cite{Comelli:2003cv,Rosati:2003yw,%
Profumo:2003hq,Catena:2004ba,Masiero:2005qu,Pallis:2005hm,Pallis:2005bb}.
We extend this work by analyzing the precision to which the LHC and
ILC can {\em simultaneously} probe dark matter and dark energy. While
colliders have traditionally been viewed as tools for discriminating
among different {\em particle physics} models, we shall demonstrate
that they are also capable of achieving sufficient precision on the
effective field theory parameters to discern the difference between
different {\em cosmological} models, e.g. the standard scenario and
the kination domination scenario.  The anticipated collider
data will allow us to probe in the laboratory a new era in
cosmological history, which may lead to significant deviations from
the standard cosmological picture
\cite{Barenboim:2006rx,Barenboim:2006jh,Drees:2007kk,Brhlik:2000dm}.

The order of presentation will be as follows.  We begin by providing an
overview of the issue of experimental tests of the WIMP hypothesis in
Section~\ref{sec:motivation}.  In Section~\ref{sec:cosmo}, we review
the effect of kination-dominated quintessence on the calculation of the
WIMP thermal relic abundance, and provide a simple map between the
kination scenario and the standard cosmological scenario.  In
Section~\ref{sec:numerics} we then
investigate the capabilities of the LHC and ILC in pinning down both the
dark matter and quintessence parameters.
Following recent ILC studies, we use study points defined within the
``mSUGRA'' or ``cMSSM'' supersymmetric framework, where the superpartner
masses are assumed to unify at the grand unification scale.  Our study
points are analogous to the set of four LCC study points (see e.g.
\cite{Baltz:2006fm}) which were chosen to represent the four ``good''
dark matter regions of the mSUGRA model. We reserve
Section~\ref{sec:conclusions} for our discussion and conclusions.

\section{Colliders as dark matter and dark energy probes}
\label{sec:motivation}

We begin by discussing in general terms how high energy colliders 
can test the WIMP hypothesis. The physics process which determines 
the present  dark matter relic anbundance, is the dark matter 
annihilation into {\em all} possible final states with SM particles. 
Once we know the total annihilation cross-section $\sigma_A$ of any 
given WIMP candidate, we can straightforwardly predict its relic abundance. 
Therefore, we can test the WIMP hypothesis by studying the properties of the 
WIMP candidates, and consequently, trying to constrain their annihilation cross-section $\sigma_A$. 

This is clearly a challenging exercise since colliders are not 
recreating the process of dark matter annihilation per se. Instead, 
given sufficient energy, colliders would produce WIMP dark matter particles, 
either directly or indirectly (in the decays of other particles). 
If those signals can be identified over the SM backgrounds, they can be studied
in order to determine the properties of the dark matter particle, most notably, its mass 
and interaction strength to the different SM particles. 
It is in this sense that colliders are helpful in probing the dark matter.

However, it is clear that the extrapolation from collider data to a
prediction about the dark matter annihilation cross-section $\sigma_A$
is rather challenging. The main problem is that the relic abundance is
determined by a multitude of final states, while at colliders we
typically observe the dark matter particles only in a limited number of
exclusive final states. It is unrealistic to expect that colliders will
be able to measure every single dark matter coupling, and the best one
could hope for is that the colliders will be able to measure the largest
(and therefore most relevant) couplings with some precision, while
placing limits on the remaining (hopefully smaller and less relevant)
couplings. Obviously, the way this is done is by either observing a dark
matter signal in a specific channel, which would provide a measurement
of the corresponding coupling, or failing to observe a signal in a
specific channel which would provide an upper limit on the corresponding
coupling. Then, by adding the results for all possible annihilation
final states, one would obtain both a lower limit $\sigma_A^{min}$ and
an upper limit $\sigma_A^{max}$ on the {\em total} annihilation
cross-section $\sigma_A$.  
Assuming standard cosmology, these limits would correspondingly
translate into an upper and lower limit on the WIMP relic density
$\Omega_\chi$. From the preceding discussion it is clear that
$\sigma_A^{min}$ (and the corresponding upper limit on $\Omega_\chi$) is
derived from collider information from channels which {\em have been
observed} while $\sigma_A^{max}$ (and the lower limit on $\Omega_\chi$)
is derived from collider information from channels which have {\em not}
been observed. It is therefore also clear that the collider information
about $\sigma_A^{min}$ is more robust than the information regarding
$\sigma_A^{max}$. First, the presence of additional annihilation
channels into non-SM final states would invalidate a bound on
$\sigma_A^{max}$ which had been derived strictly within the SM. Second,
arriving at a meaningful bound on $\sigma_A^{max}$ is {\em
experimentally} quite challenging: the unobserved channels typically
outnumber the observed ones, and also, the unobserved channels tend to
be plagued with larger backgrounds and, correspondingly, larger
uncertainties -- after all, the large backgrounds are often the reason
those channels have not been observed in the first place.

With all this in mind, let us now turn our attention to some specific
scenarios.  For any given WIMP dark matter candidate cosmology data
provides a preferred value $\sigma_A^{exp}$ for its {\em total}
annihilation cross-section. This value depends mostly on whether the
dark matter particle is an $s$-wave or $p$-wave annihilator, and is
almost insensitive to the dark matter particle spin
\cite{Birkedal:2004xn}.  The test of dark matter at colliders is being
done by comparing $\sigma_A^{exp}$ to the derived limits on
$\sigma_A^{min}$ and $\sigma_A^{max}$ as described above. Generally
speaking, there are three possible outcomes of this test, which we shall
now consider in turn.
\begin{enumerate}
\item The test will be deemed successful if it turns out that
\begin{equation}
\sigma_A^{min} < \sigma_A^{exp} < \sigma_A^{max}.
\label{outcomeA}
\end{equation}
Under those circumstances, the dark matter particle discovered 
at colliders can solely account for all of the dark matter in the universe,
{\it i.e.}, there is no need for another independent dark matter candidate.
This is the scenario which has attracted the most attention in the literature.
The gap between $\sigma_A^{min}$ and $\sigma_A^{max}$ is indicative 
of the precision with which colliders can test the WIMP dark matter hypothesis,
and with the availability of a next generation lepton collider, typically the 
amount of wiggle room between $\sigma_A^{min}$ and $\sigma_A^{max}$ 
is reduced to the order of the current uncertainty in $\sigma_A^{exp}$. 
If this turns out to be the case, we will have a triumphant 
confirmation of the WIMP dark matter hypothesis, although one could 
always find some caveats.\footnote{For example, it would be very interesting to 
see how one could reconcile such a successful test of dark matter at colliders
with a possible experimental direct detection of dark 
matter in a mass range different from the mass range found at colliders.}

Our major point here is that even though this case has been most widely 
considered in the literature, one should still pay proper attention to the 
other two possible outcomes below, especially since they would clearly indicate 
the presence of some additional new physics.
\item It is also quite possible that the collider test of dark matter will reveal that
\begin{equation}
\sigma_A^{min} < \sigma_A^{max} < \sigma_A^{exp}.
\label{outcomeB}
\end{equation}
Under the conventional interpretation, the dark matter candidate would
then overclose the universe and cannot constitute the present-day
cosmological dark matter.  This conclusion is also subject to caveats
either on the astrophysics or particle physics side.  First,
nonstandard cosmological evolution will change the picture - for
example late inflation may dilute the dark matter density so that the
particle observed at colliders is indeed the dark matter.  This scenario
can be strengthened by the observation of a direct detection signal of
dark matter with mass in the range observed at colliders. The information 
obtained at colliders will then provide invaluable insights
into early universe cosmology.  Alternatively, one may look for particle physics
resolutions of the puzzle.  One possible explanation 
is that the WIMP dark matter candidate found at colliders is
metastable and decays post-freeze out to a lighter particle 
with only gravitational interactions (superWIMP), 
which is the real dark matter \cite{Feng:2003xh,Feng:2003uy}. 
The mass ratio of the WIMP and the
superWIMP can be adjusted to compensate for the overclosure of the
universe and obtain the proper relic density today, although
significant constraints arise from BBN and large scale structures. 
Colliders are very useful in providing information, such as the mass 
of the dark matter superWIMP, which cannot be obtained by any other means.  
Indeed, direct and indirect detection experiments are
bound to fail in their searches for purely gravitationally interacting
particle dark matter.  Another possible particle physics explanation 
is to invalidate the bound $\sigma_A^{max} < \sigma_A^{exp}$. For example,
additional invisible annihilation channels not easily revealed at colliders 
would push the bound on $\sigma_A^{max}$ higher. 
In summary, outcome (\ref{outcomeB}) requires either a reconcilation of the collider results by modifying standard cosmology
and/or postulating new physics, or an invalidation of the colllider results
by postulating new physics. New physics is thus expected either
on the cosmology or particle physics side. 

\item The third possible outcome of the dark matter test at colliders is that
\begin{equation} 
\sigma_A^{exp} < \sigma_A^{min} < \sigma_A^{max}.
\label{outcomeC} 
\end{equation} 
The conventional interpretation of this
outcome is that the dark matter particle observed at colliders is
not the only component of the dark matter in the universe, and one must
look for another, yet unobserved, dark matter particle. 
Here again one may look for ways to circumvent this
conclusion. On the astrophysics side, a nonstandard cosmological history
which leads to an enhancement of the relic abundance of the dark matter
candidate, as is provided by the kination-dominated scenarios considered
in this paper, can naturally accommodate this apparently unsuccessful
dark matter particle candidate. On the particle physics side, one would
have to invent another, independent dark matter candidate, which would
provide the missing component of the dark matter. While this is possible
in principle, it requires
nonminimal model building - the presence of multiple dark matter
candidates would require the existence of multiple conserved symmetries
to ensure separately the stability of these particles on cosmological
scales. Just as before, one may think of ways to invalidate the
problematic bound $\sigma_A^{exp}<\sigma_A^{min}$. However, and this is
the major difference between outcomes (\ref{outcomeB}) and
(\ref{outcomeC}) from a particle physicist's perspective, it is clearly
almost impossible to invalidate the lower bound on the annihilation
cross-section, since, as elaborated above, it is derived from processes
which have already been firmly observed at colliders, so any channels
which may have been missed, would only increase $\sigma_A^{min}$ and
thus make the discrepancy worse.  The outcome (\ref{outcomeC}) is also
very exciting for an astrophysicist, since, in the absence of
alternative particle dark matter candidates, it would provide a direct
indication of a cosmological relic abundance enhancement mechanism (such
as kination domination) and give reasons to hunt for further correlated
astrophysical/cosmological signatures
\cite{Chung:2007vz,Profumo:2003hq}. 
\end{enumerate}

The above discussion underscores the potential importance of 
high energy colliders for the understanding of our universe.
Colliders have traditionally been viewed as tools for
{\em proving} that the WIMP particle indeed constitutes the 
dark matter. Now we also see that a potential unexpected outcome of the 
collider tests of the WIMP dark matter hypothesis may provide 
precious insights into early universe cosmology.

At the same time, one should not undervalue the potential significance
of experiments dedicated to direct dark matter searches. A positive
signal in any one of them would have several important
implications. First and foremost, it would mean that the dark matter
is real and would eliminate simple alternative explanations such as
MOND or modified gravity.  Of course, modified gravity can still play
a role in the dark matter story in the second and third possible
outcomes of the dark matter test at colliders stated above (see
e.g. \cite{Catena:2004ba,Lahanas:2006xv}), but particle dark matter
must play a significant role in gravitational clustering and galaxy
formation.  Second, it would boost the WIMP dark matter hypothesis, as
it can (depending on the inferred relic density) rule out particles
with purely gravitational interactions as a significant component of
the dark matter halo
(e.g. \cite{Chung:1998zb,Chung:2004nh,Feng:2003xh,Feng:2003uy}).
Finally, as emphasized in
\cite{Bourjaily:2005ax,Baltz:2006fm,Drees:2007hr}, direct detection
experiments will provide an important piece of the dark matter puzzle
-- an independent estimate of the mass of the dark matter particle,
which can then be contrasted with the analogous measurement at
colliders.  In the spirit of our earlier discussion, here again we can
consider three possible outcomes of this comparison:
\begin{enumerate}
\item It may turn out that the WIMP candidate found at colliders is
{\em lighter} than the directly detected dark matter particle. This
scenario would lead to several interesting possible
interpretations. For example, the assumption that the dark matter is
made up of a single component would imply that the particle found at
colliders cannot be a dark matter particle and would have to decay
outside the detector.  Alternatively, the dark matter may be made up
of several components, allowing for the coexistence of several 
particles as stable dark matter candidates.  The collider particle may
even be the dominant component if its direct detection rates are
suppressed. In any case, the direct detection results
would then provide a rough target for the next energy scale which
would need to be reached and probed by collider experiments.

\item Conversely, it may turn out that the WIMP candidate found at
colliders is {\em heavier} than the directly detected dark matter
particle.  Within the single component dark matter scenario, the
most natural explanation of the discrepancy would be that the collider
WIMP decays invisibly to the dark matter particle. Then, it would be
interesting to go back and scour the collider data for events where
the lighter dark matter particle can directly manifest itself.

\item Finally, the collider and astroparticle mass determinations may
turn out to be in agreement, which would point towards a single WIMP
dark matter component. The collider measurements of the
dark matter particle properties can then be used to reconstruct the WIMP
annhilation rate in the early universe. The outcome of this exercise
is extremely interesting. For example, either $\sigma_A^{max} <
\sigma_A^{exp}$ or $\sigma_A^{exp}<\sigma_A^{min}$ would necessarily
require nonstandard cosmology. If $\sigma_A^{max} < \sigma_A^{exp}$,
the universe appears overclosed and some mechanism of late entropy
production (such as a phase transition or a late particle decay) is
required.  If, on the other hand, $\sigma_A^{exp}<\sigma_A^{min}$, the
dark matter abundance would require some kind of a boost, such as a
period of kination domination as will be discussed in this paper.
\end{enumerate}

To summarize, the outcome of experimental tests of the WIMP hypothesis
will provide significant insights into new physics.  Although most
studies of this issue focus on scenarios in which standard cosmology
holds and the thermal WIMP discovered at colliders is the dominant
component of the cold dark matter, such that
$\sigma_A^{min}<\sigma_A^{exp}<\sigma_A^{max}$, alternative outcomes
also warrant careful consideration.  Such alternative outcomes can have
important implications for new particle physics, such as the possibility
of several distinct constituents of the cold dark matter, and/or
cosmology, such as the possiblity of cosmological enhancement or
dilution mechanisms for the Hubble expansion of the universe.  We focus
here on a particular cosmological enhancement mechanism for the dark
matter relic abundance due to the presence of dark energy in
the form of quintessence.  In this situation,
$\sigma_A^{exp}<\sigma_A^{min}<\sigma_A^{max}$, and yet the thermal WIMP
measured at colliders can be the dominant component of the cold dark matter.
Kination-dominated quintessence generically can be tested through a
number of correlated cosmological implications and signatures (see
\cite{Chung:2007vz} for a discussion and further details), but is also
a framework in which dark energy can make itself manifest at high energy
colliders, in striking contrast to the vast majority of known dark
energy models. We will now concentrate on this scenario and explore the
prospects for experimental tests at the LHC and ILC.

\section{Cosmology and the dark matter/dark energy connection}
\label{sec:cosmo}

Given the motivation for considering quintessence models of dark energy
which include a period of kination domination, we turn to its effects on
the dark matter relic abundance, as first discussed in
\cite{Salati:2002md}.  In this section, we provide a self-contained set
of analytic equations which can be used to map any ordinary dark matter
freeze out scenario to the dark matter freeze out scenario of kination
domination.\footnote{Here we will restrict our attention to a flat FRW
universe with $ds^{2}=dt^{2}-a^{2}(t)d\vec{x}^{2}$ governed by the
standard Einstein-Hilbert gravitational action. Planck's constant is
defined to be $M_{pl}\approx1.22\times10^{19}$GeV.}

In the usual thermal WIMP dark matter scenario (see e.g. \cite{KT}), the
dark matter $\chi$ is assumed to be initially in chemical equilibrium
with the thermal plasma that is in equilibrium with the photons. The
WIMPs remain in equilibrium as long as the annihilation reaction rate
$\Gamma_{A}\equiv n_{\chi}\langle\sigma_{A}v\rangle$ ($\sigma_{A}$ is
the cross section for reactions that change the number density $n_\chi$
and the averaging is with respect to a thermal ensemble) is much larger
than the expansion rate of the universe $H\equiv\dot{a}/a$ ($a$ is the
scale factor).  However, when $\Gamma_{A}$ falls below $H$, $n_{\chi}$
no longer tracks the equilibrium density and $n_{\chi}a^{3}$ is nearly
an adiabatic constant. This transition from equilibrium abundance
tracking to a nearly adiabatic constant behavior is known as the freeze
out transition. Typically, the energy density that governs $H$ during
the freeze out period is composed of relativistic degrees of freedom in
equilibrium with the photons.

The assumption that the dark energy is in the form of quintessence
naturally leads one to question whether the relativistic degrees of
freedom governing the expansion rate $H$ need be in chemical
equilibrium with the photons during WIMP freeze out. 
Indeed, since the coupling of the quintessence field to
ordinary matter needs to be very weak to maintain a long dynamical
time scale and evade fifth force constraints, quintessence naturally
never reaches chemical equilibrium with the photons. However, if the
quintessence field were to dominate the universe during freeze out,
its energy density must dilute faster than the energy density of the
SM relativistic degrees of freedom $\rho_{R}$ to evade the BBN
constraint on the number of relativistic degrees of freedom.  This
constraint can be expressed (at 95\% confidence level) as
\cite{Cyburt:2004yc}
\begin{equation} 
\Delta g_{*S}\lesssim 1.5 \frac{7}{4}, 
\end{equation} 
where $\Delta g_{*S}$ is the number of effective degrees of
freedom \cite{KT} other than the photons, electrons, and neutrinos
entering the total entropy density
\begin{equation}
s=\frac{2\pi^{2}}{45}(10.75+\Delta g_{*S})T_0^{3},
\end{equation}
where $T_0$ is the temperature during BBN.

As pointed out in \cite{Salati:2002md} , one way to achieve this is to
have a period of kination domination, in which the kinetic energy of the
quintessence field $\Phi$ dominates the energy density:
$\rho_{\Phi}\approx\frac{1}{2}\dot{\Phi}^{2}\gg\rho_{R}.$ In this case,
the energy density scales as
\begin{equation}
\frac{\rho_{\Phi}}{\rho_{R}}\propto\frac{1}{a^{2}}.
\end{equation} 
This indicates that if the ratio of energy density of $\Phi$ and photons at the
temperature of $1$ MeV is
\begin{equation}
\eta_{\Phi}\equiv\frac{\rho_{\Phi}}{\rho_{\gamma}}|_{T=1{\rm{MeV}}},
\label{eq:kinpar}
\end{equation}
the ratio at the time of freeze out is of the order
\begin{equation}
\frac{\rho_{\Phi}}{\rho_{\gamma}}|_{T=1 {\rm GeV}}\sim 10^{6}\eta_{\Phi}.
\end{equation} 
This indicates that
the $\Phi$ energy density dominates over that of the relativistic
degrees of freedom, since $\rho_{\Phi}\sim10^{5}\eta_{\Phi}\rho_{R}$ at the
approximate period of freeze out and 
\begin{equation}
0\leq\eta_{\Phi}\lesssim1.
\end{equation}  
Therefore, we can easily arrange
$\rho_{\Phi}$ to control the freeze out temperature $T_{K}$ in the kination dominated
scenario. (Note that one should distinguish between $\rho_{\gamma}$
which is the density of photons and $\rho_{R}$ which is the energy
density of all relativistic particles.)

Following the computational approach of \cite{Salati:2002md}, we can
easily give a map between the dark matter abundance of the standard 
scenario and that of the kination domination scenario. 
With LSP dark matter in mind, the thermal averaged annihilation 
cross section (defined to be proportional to the number changing
reaction rate) is
\begin{equation}
\langle\sigma_{A}v\rangle=\tilde{a}+\tilde{b}x,
\label{eq:crosssection}
\end{equation}
where 
\begin{equation}
x\equiv\frac{T}{m_{\chi}},
\end{equation}
$m_{\chi}$ is the mass of the WIMP, and $\tilde{a}$ ($\tilde{b}$) is the s-wave (p-wave)
contribution to the annihilation cross section. The freeze out temperature
parameter will be denoted by $x_{U}$ in the standard scenario and 
$x_K$ in the kination domination scenario. 
These quantities are defined by the equations
\begin{equation}
(\tilde{a}+\tilde{b}x_{U,K})n_{\chi}^{eq}(x_{U,K})=\frac{2}{x_{U,K}}H_{U,K}(x_{U,K}),
\end{equation}
\begin{equation}
n_{\chi}^{eq}(x)=2m_{\chi}^{3}(\frac{x}{2\pi})^{3/2}e^{-\frac{1}{x}},
\end{equation}
\begin{equation}
H_{U}(x)=\sqrt{\frac{8\pi^{3}}{90}g_{*}(x)}\frac{m_{\chi}^{2}}{M_{pl}}x^{2},
\end{equation}
\begin{equation}
H_{K}(x)=\sqrt{\frac{8\pi^{3}}{90}g_{*}(x)}\frac{m_{\chi}^{2}}{M_{pl}}x^{2}\sqrt{1+\alpha x^{2}},
\end{equation} 
where $g_{*}(x)$ is defined by the equation for the radiation energy density
\begin{equation}
\rho_{R}=\frac{\pi^{2}}{30}g_{*}(x)x^{4}m_{\chi}^{4}.
\end{equation}
In the kination domination scenario, the relic abundance today $\Omega_{K}$ can then be expressed relative to
the usual relic abundance $\Omega_U$ as
\begin{equation}
\frac{\Omega_{K}}{\Omega_{U}}=\frac{\mu_{U}}{\mu_{K}}\frac{x_{U}^{2}}{x_{K}^{2}}\sqrt{\frac{g_{*}(x_{U})}{g_{*}(x_{K})}}\sqrt{1+\alpha x_{K}^{2}}\left[\frac{\tilde{a}+\tilde{b}x_{U}}{\tilde{a}+\tilde{b}x_{K}}\right],
\label{eq:kinetictousual}
\end{equation}
\begin{equation}
\mu_{U}=1+\frac{2}{x_{U}}\left(\frac{\tilde{a}+\tilde{b}x_{U}/2}{\tilde{a}+\tilde{b}x_{U}}\right),
\end{equation}
\begin{equation}
\mu_{K}=1+\frac{2}{x_{K}}\sqrt{1+\alpha x_{K}^{2}}\left(\frac{\tilde{a}A(u)+\tilde{b}x_{K}B(u)}{\tilde{a}+\tilde{b}x_{K}}\right),
\end{equation}
\begin{equation}
A(u)=\frac{1}{u}\ln[u+\sqrt{1+u^{2}}],
\end{equation}
\begin{equation}
B(u)=\frac{\sqrt{1+u^{2}}-1}{u^{2}},
\end{equation}
\begin{equation}
u=\sqrt{\alpha}x_{K}\end{equation}
\begin{equation}
\alpha=\eta_{\Phi}\frac{m_{\chi}^{2}}{[g_{*}(T_{0})/2]T_{0}^{2}}\left[\frac{g_{*}(T_{K})}{g_{*}(T_{0})}\right].
\end{equation}
In the above formulas, a subscript $U$ denotes quantities associated
with the usual freeze out scenario, while a subscript $K$ denotes
quantities associated with the kination domination scenario. Here the
number of relativistic degrees of freedom governing the energy density
$g_{*}$ obeys the approximate relationship $g_{*}\approx g_{*S}$ (e.g.
$g_{*}(T_{K})\approx90$ and $g_{*}(T_{0})\approx10.75$), and
$T_{0}=1\textrm{ MeV}$ is the temperature relevant for BBN (as opposed
to the temperature today). 

Assuming that $\eta_{\Phi}\leq1$ is sufficiently large such that
kination dominates at freeze out, let us approximate
Eq.~(\ref{eq:kinetictousual}) in the limit of either $s$-wave ($\tilde a
>> \tilde b x_K$) or $p$-wave ($\tilde a << \tilde b x_K $) dominance to
obtain intuition for the type of enhancement obtained for kination
domination. This is useful because in any given realistic model the dark
matter candidate tends to be either a predominantly $p$-wave annihilator
(e.g. the neutralino in supersymmetry) or an $s$-wave annihilator (e.g.
the dark matter candidate in most other cases).  For a $p$-wave
annihilator we obtain
\begin{equation}
\frac{\Omega_{K}}{\Omega_{U}}\sim\frac{g_{*S}(T_{U})}{g_{*S}(T_{0})}\frac{T_{U}^{2}}{T_{K}T_{0}}\frac{\sqrt{\eta_{\Phi}}}{\sqrt{2 g_{*}(T_{U})}}.
\label{eq:approxdependence}
\end{equation}
The powers of the temperature $T$ in Eq.~(\ref{eq:approxdependence}) can be
understood as follows. Since
$\Omega_{U,K}\propto(a_{U,K})^{3}/a_{\rm{today}}^{3}$
and the volume dilution behaves as $1/a^{3}\propto T^{3}$, the volume
factors alone contribute $(T_{U}/T_{K})^{3}$.   Residual annihilations
after freeze out contribute a factor of order $T_{K}/T_{U}$
for $p$-wave dominance.
The most nontrivial aspect of the kination scenario is
that since the freeze out condition is $\langle\sigma v\rangle
n_{\chi}=m_\chi H/T$, with \begin{equation}
H_{K}\sim\sqrt{\eta_{\Phi}}\left (\frac{T_{K}}{T_{0}}\right )^{3}\frac{T_{0}^{2}}{M_{pl}}\end{equation}
(recall $H^{2}\propto\eta_{\Phi}/a^{6}$ during kination domination),
$H_{U}\sim T_{U}^{2}/M_{pl}$, and $\langle\sigma v\rangle\propto T$ ($p$-wave
dominance), there is a factor of $T_{K}/T_{0}$.  
Note that since $T_{0}/T_{U}\sim10^{-3}$ and the difference between $T_{K}$
and $T_{U}$ is only logarthmically dependent on the $\dot{\Phi}^{2}$
energy density, it is unrealistic to obtain a ratio smaller than unity in
Eq.~(\ref{eq:approxdependence}).

We similarly approximate Eq. (\ref{eq:kinetictousual}) in the limit of $s$-wave
dominance to obtain
\begin{equation}
\frac{\Omega_{K}}{\Omega_{U}}\sim\frac{g_{*S}(T_{U})}{g_{*S}(T_{0})}\frac{T_{U}}{T_{0}}\frac{\sqrt{\eta_{\Phi}}}{\log(2u)\sqrt{g_{*}(T_{U})}/2}.
\label{eq:approxsdependence}
\end{equation}
Inserting typical numbers $\{ m_{\chi}\sim100\textrm{ GeV},\textrm{
}T_{0}\sim10^{-3}\textrm{ GeV},\textrm{ }g_{*S}(T_{U})\sim
g_{*}(T_{K})\sim10^{2},\textrm{ }g_{*S}(T_{0})\approx10.75,\textrm{
}\tilde{b}=5.4\times10^{-8}\ \textrm{GeV}^{-2},\eta_{\Phi}\sim1\}$
into Eq.~(\ref{eq:approxdependence}) and
Eq.~(\ref{eq:approxsdependence}), we generically find for both cases
\begin{equation}
\frac{\Omega_{K}}{\Omega_{U}}\sim10^{3}.
\label{eq:Omegaratios}
\end{equation}
Hence, the kination scenario relic abundance is much larger than
the usual freeze out scenario relic abundance for the same microphysical
parameters governing the WIMP annihilation cross section.
\FIGURE[t]{
\epsfig{file=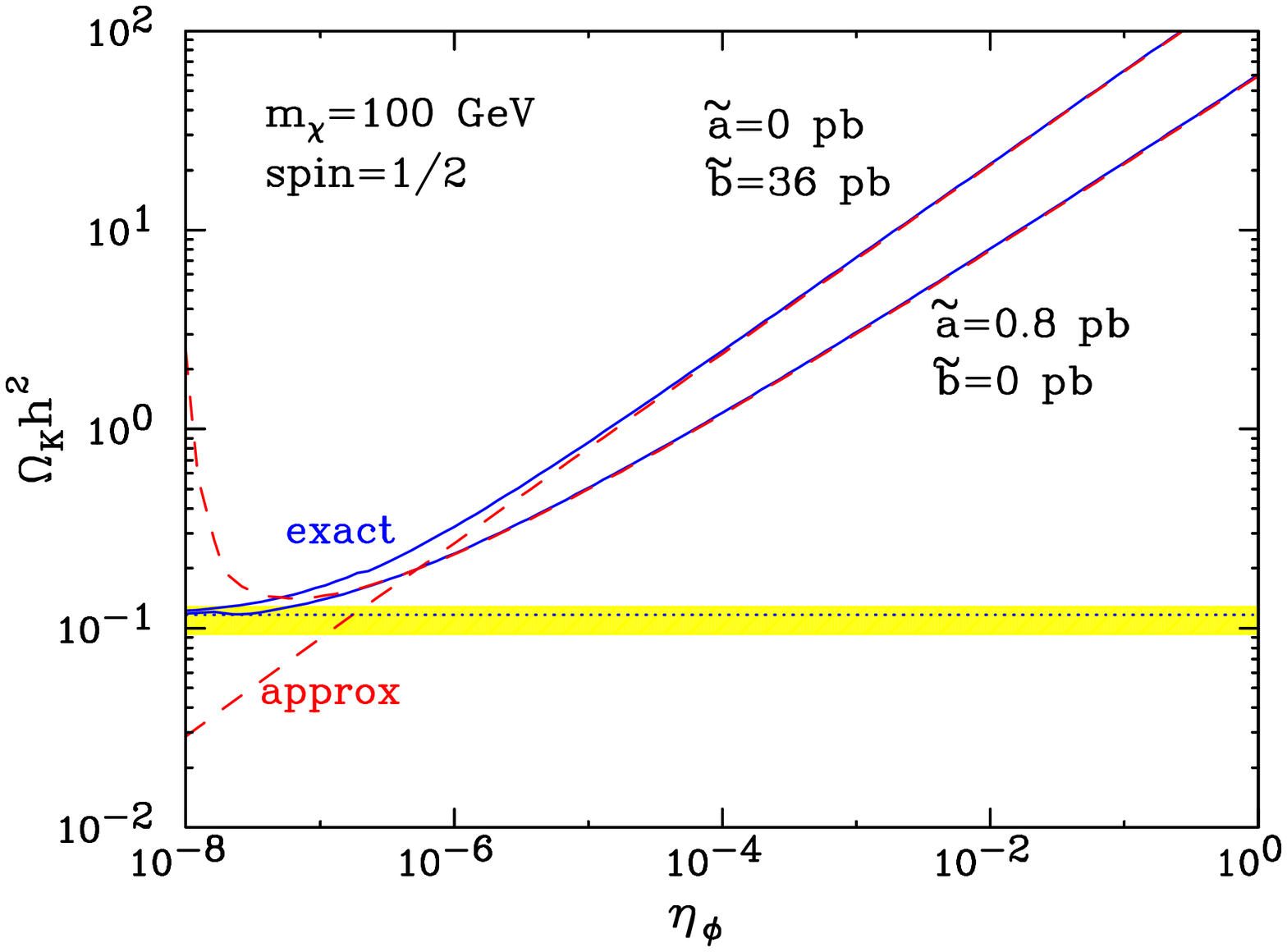,width=9.0cm}
\caption{\sl The dark matter relic abundance $\Omega_K h^2$ in the
kination domination scenario, as a function of the kination parameter
(\ref{eq:kinpar}), for the case of a pure $s$-wave annihilator ($\tilde
a=0.8$ pb, $\tilde b=0$) or a pure $p$-wave annihilator ($\tilde a=0$,
$\tilde b=36$ pb) with spin 1/2 and mass $m_\chi=100$ GeV.  The (blue)
solid lines give the exact result from Eq.~(\ref{eq:kinetictousual}),
while the (red) dashed lines correspond to the approximations
(\ref{eq:approxsdependence}) and (\ref{eq:approxdependence}). The
horizontal dotted line and the (yellow) shaded band denote the current
central value and 2$\sigma$ range for the experimental determination of
the dark matter relic abundance.}
\label{fig:approx}}
This is illustrated in Figure~\ref{fig:approx}, which shows the dark
matter relic abundance $\Omega_K h^2$ in the kination domination
scenario, as a function of the kination parameter (\ref{eq:kinpar}), for
the case of a pure $s$-wave annihilator ($\tilde a=0.8$ pb, $\tilde
b=0$) or a pure $p$-wave annihilator ($\tilde a=0$, $\tilde b=36$ pb).
We assume that the dark matter particle is a fermion with spin 1/2 and
mass $m_\chi=100$ GeV.  The values for its annihilation cross-sections
are chosen so that in the usual scenario ($\eta_\Phi=0$) this particle
would make up all of the dark matter, in agreement with the experimental
determination of the dark matter relic abundance by WMAP and SDSS
($\Omega_\chi h^2 = 0.111_{-0.015}^{+0.011}$ at 2$\sigma$
\cite{Spergel:2006hy}, as indicated by the horizontal dotted line and
the (yellow) shaded band). The (blue) solid lines give the exact result
from Eq.~(\ref{eq:kinetictousual}), while the (red) dashed lines
correspond to the approximations (\ref{eq:approxsdependence}) and
(\ref{eq:approxdependence}). We see that the approximate expressions
work quite well, down to $\eta_\Phi\sim 10^{-7}$ for $s$-wave
annihilators, and $\eta_\Phi\sim 10^{-5}$ for $p$-wave annihilators.
While the approximations (\ref{eq:approxsdependence}) and
(\ref{eq:approxdependence}) are useful for understanding the scaling of
the relic density in the presence of quintessence, in our numerical
results below we shall always make use of the exact expression
(\ref{eq:kinetictousual}).  From Fig.~\ref{fig:approx} we also see that,
in agreement with the naive expectation (\ref{eq:Omegaratios}), for
$\eta_\Phi\sim 1$ and at this value of $m_\chi=100$ GeV, kination
domination provides an enhancement by about three orders of magnitude of
the dark matter thermal relic density.
\FIGURE[t]{
\epsfig{file=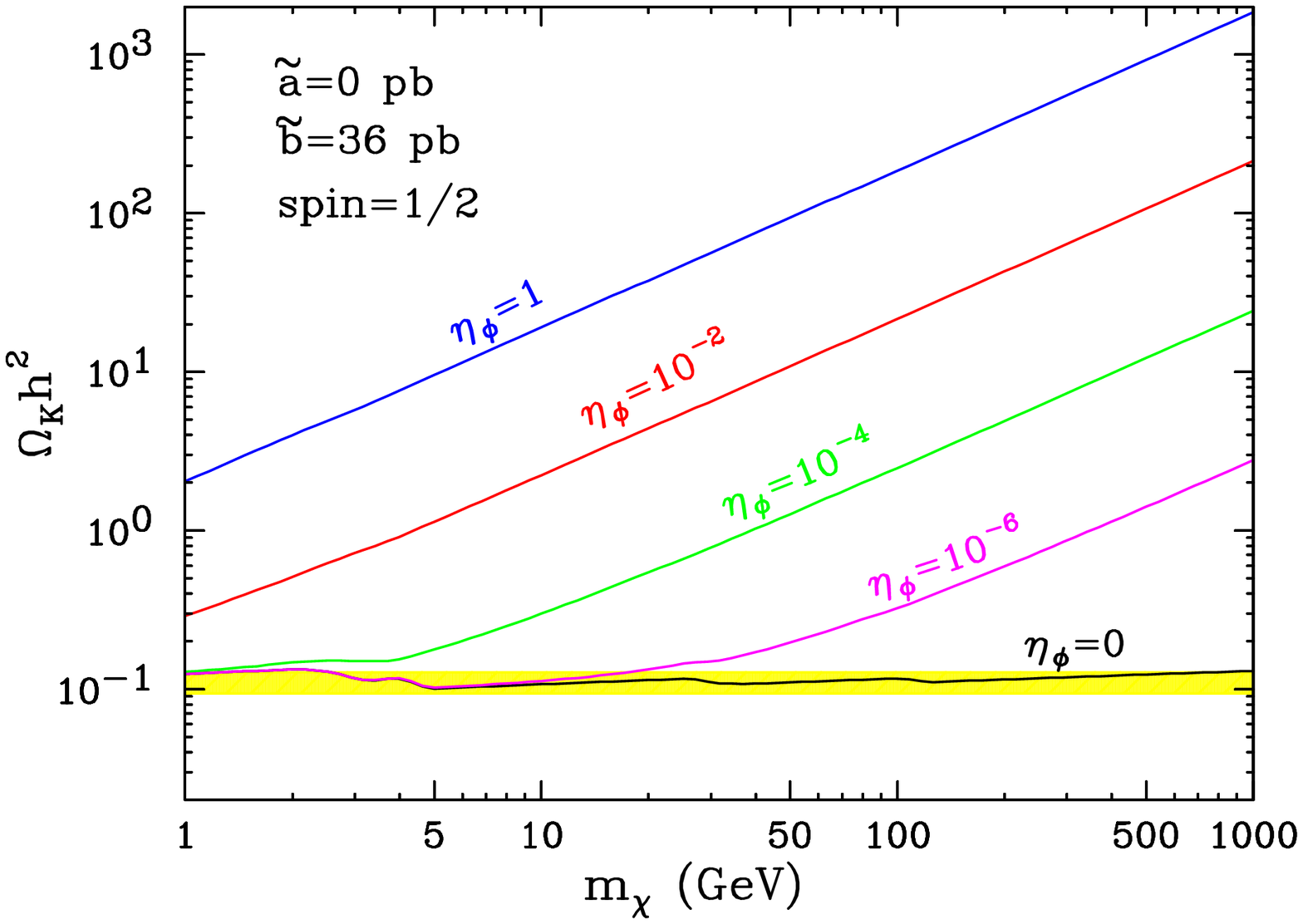,width=9.0cm}
\caption{\sl The dark matter relic abundance $\Omega_K h^2$ in the
kination domination scenario, as a function of the mass $m_\chi$ of the
dark matter particle, for the case of a pure $p$-wave annihilator
($\tilde a=0$, $\tilde b=36$ pb) and different values of the kination
parameter $\eta_\Phi$. The horizontal (yellow) shaded band denotes the
current 2$\sigma$ range for the experimental determination of the dark
matter relic abundance.}
\label{fig:omegaK}}
The kination enhancement of the dark matter relic density can be even
more pronounced, if the mass $m_\chi$ of the dark matter particle is
larger, as evidenced in Figure~\ref{fig:omegaK}. There we plot the dark
matter relic abundance $\Omega_K h^2$ in the kination domination
scenario, this time as a function of the mass $m_\chi$ of the dark
matter particle, for the case of a pure $p$-wave annihilator ($\tilde
a=0$, $\tilde b=36$ pb) and different values of the kination parameter
$\eta_\Phi$. As in Fig.~\ref{fig:approx}, the horizontal (yellow) shaded
band denotes the current 2$\sigma$ range for the experimental
determination of the dark matter relic abundance. We see for example,
that if the dark matter matter particle had a mass of order 1 TeV,
kination may provide an enhancement of its relic density of up to 4
orders of magnitude!

\FIGURE[t]{
\epsfig{file=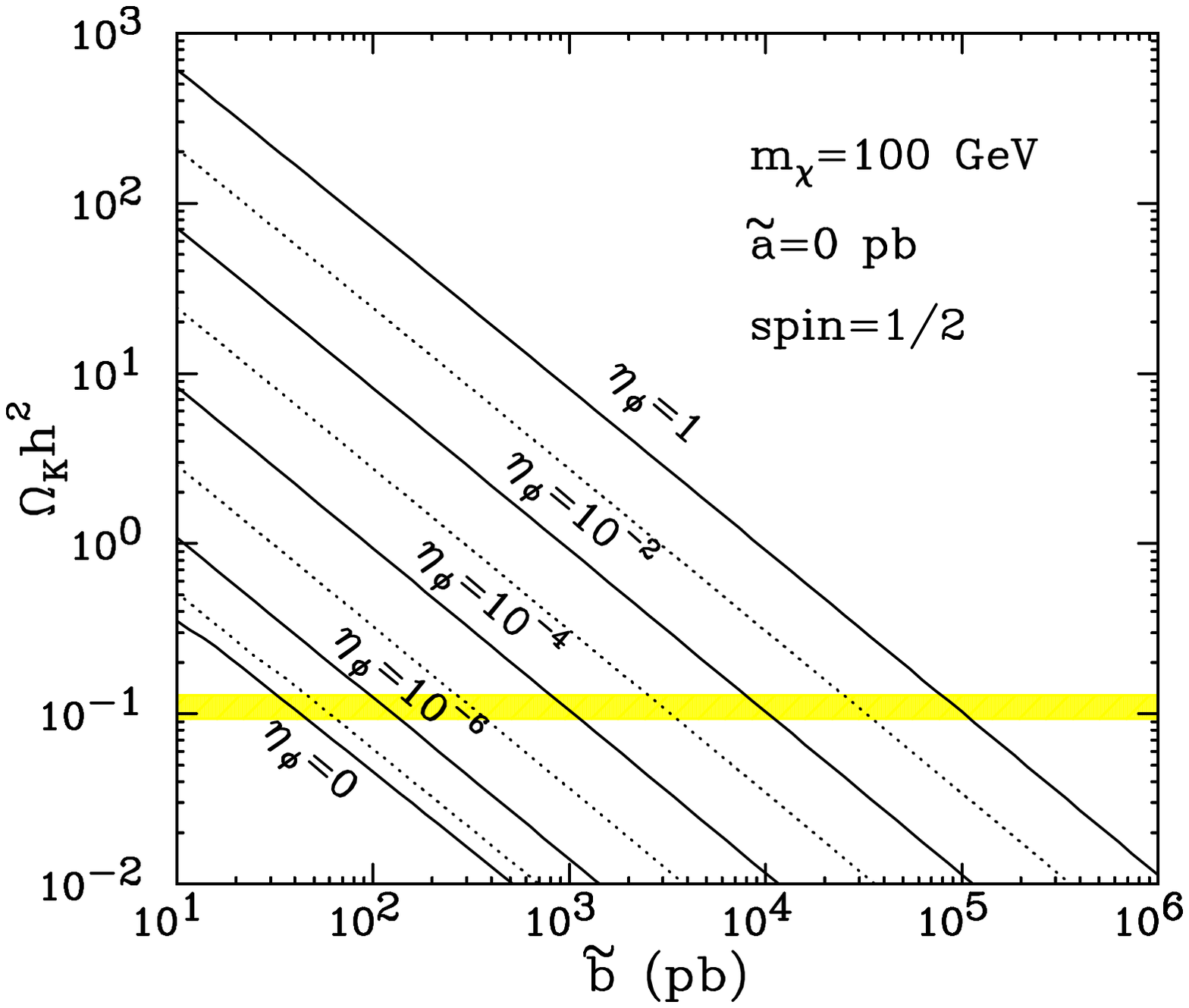,width=10.0cm}
\caption{\sl The dark matter relic abundance $\Omega_K h^2$ in the
kination domination scenario, as a function of the annihilation
cross-section $\tilde b$ of the dark matter particle, for the case of a
pure $p$-wave annihilator ($\tilde a=0$) of spin $1/2$ and mass
$m_\chi=100$ GeV, for different values of the kination parameter
$\eta_\Phi$.  The solid lines are plotted for $\eta_\Phi$ values as
labelled on the plot (every other decade), while the dotted lines
correspond to intermediate decades.  The horizontal (yellow) shaded band
denotes the current 2$\sigma$ range for the experimental determination
of the dark matter relic abundance.}
\label{fig:omegaK2}}

Given these results, let us return to the issue of testing the WIMP
hypothesis at colliders as discussed in Section~\ref{sec:motivation}. To
make the discussion more concrete, consider the following scenario.
Suppose that colliders have found a WIMP of mass 100 GeV whose spin was
measured to be 1/2. The lightest neutralino in supersymmetry is a
standard such example. Furthermore, because of its Majorana nature, the
neutralino typically annihilates predominantly in a p-wave, so $\tilde a
<< \tilde b$. The measurements of the neutralino couplings and the
superpartner mass spectrum can then be translated into a bound on the
annihilation cross-section $\sigma_A$, in this case its $\tilde b$
component. Given the bounds on $\tilde b$, one can then reconstruct the
allowed range for the WIMP relic abundance, both with and without the
effects of kination domination, as shown in Fig.~\ref{fig:omegaK2}.
There we plot the dark matter relic abundance $\Omega_K h^2$ in the
kination domination scenario, as a function of the annihilation
cross-section $\tilde b$ of the dark matter particle, for different
values of the kination parameter $\eta_\Phi$. As in
Figs.~\ref{fig:approx} and \ref{fig:omegaK}, the horizontal (yellow)
shaded band denotes the current 2$\sigma$ range for the experimental
determination of the dark matter relic abundance. The prediction of
standard cosmology (no kination domination) is given by the
$\eta_\Phi=0$ line. 

As we alluded to earlier, the case which is especially interesting is
the one where $\tilde b$ is determined to be {\em too large}, e.g.
larger than, say, 100 pb, corresponding to outcome~(\ref{outcomeC}) in
our discussion in the previous section.\footnote{Indeed, other 
outcomes of the collider tests of the WIMP hypothesis, Eqs.~(\ref{outcomeA}) and (\ref{outcomeB}), would strongly disfavor
the kination-dominated quintessence dark matter scenario and place an
upper bound on the quintessence parameter $\eta_\Phi$. Of course,
quintessence die-hards may still argue that there are extra unseen
annihilation channels into non-SM states. The corresponding increase in
$\sigma_A$ can be compensated by a nonzero value of $\eta_\Phi$. While
this is possible in principle, it would seem coincidental and rather
fine-tuned.}  Such a result can be interpreted in one of two ways.
First, assuming standard cosmology, one can read off from the
$\eta_\Phi=0$ line the maximum fraction that such a WIMP can contribute
to the dark matter budget of the Universe. Alternatively, assuming that
the WIMP makes up 100\% of the dark matter in the universe and following
the horizontal shaded band, the lower bound on $\tilde b$ would imply a
lower bound on $\eta_\Phi$, {\it i.e.}, a minimum value for the kinetic energy
contribution of the quintessence field $\Phi$ to the total energy budget
in the early universe. We would like to emphasize once again that the
lower bound on $\tilde b$ would tend to be rather robust and difficult
to invalidate - it would be derived based on some observed channel which
would guarantee a minimum value for the WIMP annihilation rate.

\section{Numerical Results}
\label{sec:numerics}

In the previous section we have seen the possible interplay between dark
matter and dark energy in the early universe. By now it is well
appreciated that high-energy colliders offer the unique opportunity of
creating and studying dark matter in the lab. Therefore, if there is a
cosmological connection between dark matter and dark energy, high-energy
colliders can also shed light on the nature of dark energy itself. We
now analyze specific WIMP scenarios in the context of low energy
supersymmetry, for which the LSP dark matter candidate is the lightest
neutralino, to ascertain to what extent forthcoming and future collider
experiments can probe dark energy and dark matter within this class of
quintessence models of dark energy.

\subsection{Choice of benchmark models}
Hence, in the remainder of this section we shall use several study cases
to investigate the capabilities of the LHC and ILC in determining the
relevant dark matter and dark energy parameters.  It has become
customary to perform such studies, using specific ``study points'' in
the parameter space of simple models. In particular, in case of
supersymmetry, the model of choice has been ``mSUGRA'', as it has rather
few input parameters. Three of them fix the values of the soft
supersymmetry breaking parameters at the unification scale: the
universal scalar mass $m_0$, the universal gaugino mass $M_{1/2}$, and
the common trilinear term $A_0$. The remaining two mSUGRA model
parameters are $\tan\beta$, the ratio of the Higgs vacuum expectation
values, and the sign of the $\mu$ term in the superpotential. While the
mSUGRA model is not representative of {\em every} possible incarnation
of supersymmetry, it is nevertheless sufficiently general to exhibit
four different regions in parameter space with a good dark matter
candidate. Correspondingly, all of the recent sets of benchmark points
suggested in the literature
\cite{Battaglia:2001zp,Allanach:2002nj,Battaglia:2003ab} have zeroed in
on those regions, and proposed study points where the supersymmetric
dark matter candidate (neutralino) makes up all of the dark matter in
the universe.

In our case, we would like to include in our analysis the effect of
kination, which, as we have already seen, tends to enhance the nominal
prediction of the dark matter relic density. Therefore, the usual sets
of mSUGRA benchmark points are ill suited for our purposes, if our goal
is to provide a {\em measurement} of the kination parameter at
colliders. Indeed, if the nominal calculation already yields a
prediction in exact agreement with experiment, kination will then make
things worse, and we can at best only place an upper limit on
$\eta_\Phi$. We have therefore chosen to modify the original set of LCC
benchmark points of \cite{Baltz:2006fm}, so that the nominal calculation
would yield a value for the relic density which is insufficient to
explain all of the dark matter in the universe. Barring the existence of
another, undiscovered yet dark matter candidate, the collider results
could then be interpreted as measurements of $\eta_\Phi$ and would
provide a non-trivial link between dark matter and dark energy. The
values for the original LCC benchmark points (LCC1-LCC4) and our
modified versions (LCC1$'$-LCC4$'$) are listed in
Table~\ref{tab:points}. The motivation behind each choice will be
discussed in the following subsections. At this point we would only
mention that we have tried to only minimally deviate from the original
LCC points, and that in all but one case our point differs from its LCC
counterpart in the value of a single mSUGRA parameter.

Previous studies have already estimated the expected accuracy with which
high energy physics experiments can pinpoint the supersymmetry
parameters at the original LCC benchmark points. One should keep in
mind, however, that those estimates are only ``best guesses'' so far,
and one would have a better idea of the actual precision only after the
colliders have been operational for some time, which would allow for
better understanding of the systematic uncertainties. Nevertheless, we
can reliably estimate the collider precision for our LCC$'$ benchmarks.
The LCC$'$ benchmark points yield rather large values for the WIMP
annihilation rate, which is typically due to the dominance of a single
channel: either an $s$-channel resonance or a coannihilation channel.
Under those circumstances, it is important to know with great precision
only the masses of the particles involved in the dominant channel, while
the rest of the spectrum may remain rather uncertain, as long as it does
not contribute significantly to the WIMP annihilation rate.  In what
follows, we shall therefore base our prediction for the collider
precision in measuring the cosmological parameters, on the expected
precision in measuring the masses of the particles entering the dominant
annihilation channel.  A more sophisticated analysis using the full
information about the SUSY spectrum and utilizing Markov chain
probabilistic techniques along the lines of \cite{Baltz:2006fm} is
beyond the scope of our paper, and will not, we believe, significantly
change our conclusions.

%%%% DO NOT CHANGE BELOW THIS LINE

\TABULAR{||c||r|r||r|r||r|r||r|r||}
{\hline\hline	
                 & LCC1 & LCC1$'$ & LCC2 & LCC2$'$ & LCC3 & LCC3$'$ & LCC4 & LCC4$'$ \\ \hline \hline
$m_0$            &  100 &   100 & 3280 & 3260  &  213 &   205 &  380 &  950  \\
$M_{1/2}$        &  250 &   150 &  300 &  300  &  360 &   360 &  420 &  420  \\
$\tan\beta$      &   10 &    10 &   10 &   10  &   40 &    40 &   53 &   50  \\ 
$A_0$            & -100 &  -100 &    0 &    0  &    0 &     0 &    0 &    0  \\
${\rm sign}(\mu)$&   +1 &    +1 &   +1 &   +1  &   +1 &    +1 &   +1 &   -1  \\
$m_t$            &  175 &   175 &  175 &  175  &  175 &   175 &  178 &  178  \\ \hline\hline
$\Omega_U h^2$   & 0.193&0.00127& 0.106& 0.0363& 0.121&0.0329 & 0.104& 0.0244\\ \hline\hline
$\tilde\chi^0_1$ &  95.5&   53.6& 107.7&   92.1& 142.6& 142.6 & 169.1& 171.8 \\ 	
$\tilde\chi^0_2$ & 181.6&   98.6& 166.3&  147.2& 274.2& 274.1 & 327.1& 335.7 \\
$\tilde\chi^0_3$ & 356.6&  232.3& 190.0&  147.9& 462.8& 462.8 & 539.7& 552.8 \\	
$\tilde\chi^0_4$ & 375.6&  256.2& 294.2&  286.2& 478.0& 478.0 & 553.0& 563.0 \\
$\tilde\chi^+_1$ & 181.6&   97.5& 159.4&  124.6& 274.5& 274.4 & 327.5& 335.1 \\ 	
$\tilde\chi^+_2$ & 374.7&  255.3& 286.6&  278.3& 478.2& 478.2 & 553.2& 563.1 \\ \hline
$\tilde     e_R$ & 143.1&  122.2&3277.3& 3357.1& 254.9& 248.3 & 412.4& 962.1 \\ 	
$\tilde     e_L$ & 204.6&  151.2&3280.1& 3359.8& 328.9& 323.9 & 477.2& 990.3 \\ 	
$\tilde   \nu_e$ & 186.2&  127.7&3276.4& 3356.1& 316.3& 311.0 & 468.2& 985.8 \\
$\tilde  \tau_1$ & 134.5&  113.2&3251.6& 3330.8& 154.9& 147.3 & 195.5& 734.0 \\
$\tilde  \tau_2$ & 207.6&  155.7&3267.7& 3346.9& 333.3& 329.3 & 441.7& 893.3 \\ 
$\tilde\nu_\tau$ & 185.3&  126.8&3263.8& 3343.0& 297.6& 292.9 & 409.2& 885.7 \\ \hline
$h$              & 113.8&  108.7& 118.7&  118.8& 116.7& 116.7 & 118.9& 118.6 \\ 	
$A$              & 394.4&  255.1&3242.2& 3318.7& 429.5& 427.6 & 419.4& 352.5 \\ \hline
$\tilde     u_R$ & 547.8&  352.6&3311.0& 3389.3& 780.2& 778.1 & 943.5&1274.9 \\ 	
$\tilde     u_L$ & 564.4&  360.4&3301.3& 3380.1& 805.0& 802.9 & 971.3&1292.6 \\ 	
$\tilde     d_R$ & 547.6&  354.4&3313.4& 3391.3& 778.5& 776.3 & 941.2&1273.6 \\
$\tilde     d_L$ & 570.4&  369.7&3302.3& 3381.0& 809.3& 807.2 & 974.8&1295.1 \\
$\tilde     t_1$ & 400.9&  237.9&1976.1& 2023.0& 602.5& 601.4 & 715.6& 872.9 \\
$\tilde     t_2$ & 577.7&  407.8&2719.9& 2783.8& 764.9& 763.8 & 875.4&1009.0 \\
$\tilde     b_1$ & 514.3&  327.4&2709.9& 2773.6& 691.0& 689.5 & 795.1& 950.8 \\ 
$\tilde     b_2$ & 538.7&  348.8&3241.3& 3318.3& 743.0& 741.5 & 861.8&1011.6 \\ \hline
$\tilde g$       & 611.2&  386.2& 850.1&  852.5& 856.2& 855.9 & 993.0&1027.5 \\ \hline
\hline}{\label{tab:points} 
mSUGRA parameter sets for the four LCC study points and their variations used in this study.
We also show the relic density and the superpartner spectrum (the masses are listed in units of GeV) 
predicted at each point
as calculated with micrOMEGAs \cite{Belanger:2001fz,Belanger:2004yn,Belanger:2006is}
and {\tt ISAJET} version 7.69 \cite{Paige:2003mg}.}

\subsection{LCC1$'$: a study point in the bulk region}
\label{sec:bulk}

In the remainder of this section we shall present our results for the 
expected accuracy in determining the dark matter properties at colliders.
We shall first discuss the physics and the relevant measurements at each 
one of our benchmark points LCC1$'$-LCC4$'$. Our predictions for the 
expected allowed range in $\Omega_K h^2$ and $\eta_\Phi$
as determined by the LHC and ILC for each study point 
will be summarized in Fig.~\ref{fig:summary}.
The superpartner mass spectra for the four original LCC study points and 
for our modified LCC$'$ study points are listed in Table~\ref{tab:points}.

The LCC1 study point was chosen in the bulk region, where the sleptons 
are relatively light, and neutralino annihilation proceeds predominantly 
through $t$-channel right-handed slepton exchange. At LCC1, there are 
7 channels which contribute more than 1\% to the annihilation rate, 
but the three dominant channels are $\tau^+\tau^-$ (32\%) and $\mu^+\mu^-$ and 
$e^+e^-$ (at 29\% each). The remaining channels (neutrino pairs and 
$b\bar{b}$) are about 2\% each. As one can see, the right-handed sleptons 
play the most important role in mediating neutralino annihilations --
first, because they are the lightest sfermions in the spectrum, and second, 
because they have the largest hypercharge (the LSP is $97.4\%$ Bino at 
this point and its couplings are proportional to hypercharge).
Under those circumstances, in order to pinpoint the WIMP relic density,
one has to measure precisely the masses of the right handed sleptons and the LSP, 
at the same time making sure that the remaining sparticles in the spectrum 
are relatively heavy. At the LHC, the slepton masses are difficult to measure 
in direct slepton production \cite{Andreev:2004qq,Birkedal:2005cm}, 
due to the relatively small slepton production cross-section and large SM 
backgrounds. Fortunately, point LCC1 has relatively light colored superpartners 
(squarks and gluino) which can be produced abundantly at the LHC and in their 
cascade decays may yield sleptons indirectly. In particular, the decays
of $\tilde\chi^0_2$ are predominantly to right-handed sleptons (since they 
are the only sfermions lighter than $\tilde\chi^0_2$). The measurements are 
by no means trivial, since one would have to extract the slepton and neutralino 
masses from a sufficient set of observed kinematic endpoints. However, 
the expectations are that one could measure the LSP mass to within 5\% and the 
neutralino-slepton mass differences to within a few GeV (for more details, 
see \cite{Baltz:2006fm}). Of course, since the sleptons at point LCC1 are 
within the kinematic reach of the ILC, their masses (as well as the LSP mass) 
can be determined at the per mil level at the ILC.

As we discussed earlier, our strategy will be to modify the original LCC 
benchmark points so that the resulting relic density is {\em too low} and can
accommodate the kination dominated quintessence scenario.
Notice that the relic density at point LCC1 is already too large\footnote{To 
some extent this is related to ensuring that the Higgs boson mass is above 
the LEP limit (we shall return to this point later on).}
(almost double the WMAP value), which appears to make our job rather difficult.
There are several ways to reduce the value of $\Omega h^2$. For example, 
one could attempt to reduce the slepton masses even further. 
However, this will not help in our case -- first, it will eventually
take us into the coannihilation region discussed later on in Section \ref{sec:coann}.
Furthermore, the value of $m_0$ at point LCC1 
is already small enough so that the slepton masses are dominated by the radiative 
corrections due to the gaugino terms in the RGE's. The slepton masses 
are therefore much more sensitive to the parameter $M_{1/2}$. This is why 
we choose to modify the LCC1 point by changing the value of $M_{1/2}$.
The resulting variation in the relic density $\Omega_U h^2$
as calculated within standard cosmology, is shown in Fig.~\ref{fig:bulk}a,
for fixed $m_0=100$ GeV, $A_0=-100$ GeV, $\tan\beta=10$, $\mu>0$ and $m_t=175$ GeV.
The horizontal (green) shaded band denotes the current 2$\sigma$ range for the 
experimental determination of the dark matter relic abundance.
The vertical (yellow) shaded band on the left is ruled out  by the negative
chargino searches at LEP. Since the parameter $M_{1/2}$ controls the values of
the gaugino masses, when it becomes too small, there will be a light (wino-like) 
chargino in the spectrum. It should be kept in mind that the LEP bound on the
chargino mass is also a function of the electron sneutrino mass $\tilde\nu_e$:
the bound is diluted in the presence of a light sneutrino 
due to destructive interference between the $Z/\gamma$ and $\tilde\nu_e$
mediated diagrams of chargino production. 
When moving in the opposite direction -- increasing $M_{1/2}$ -- we encounter
a region where the lightest slepton ($\tilde\tau_1$) 
becomes increasingly lighter relative to the lightest neutralino ($\tilde\chi^0_1$)
and eventually becomes the LSP in the region denoted by the vertical (yellow) 
shaded band on the right.

\FIGURE[t]{
\epsfig{file=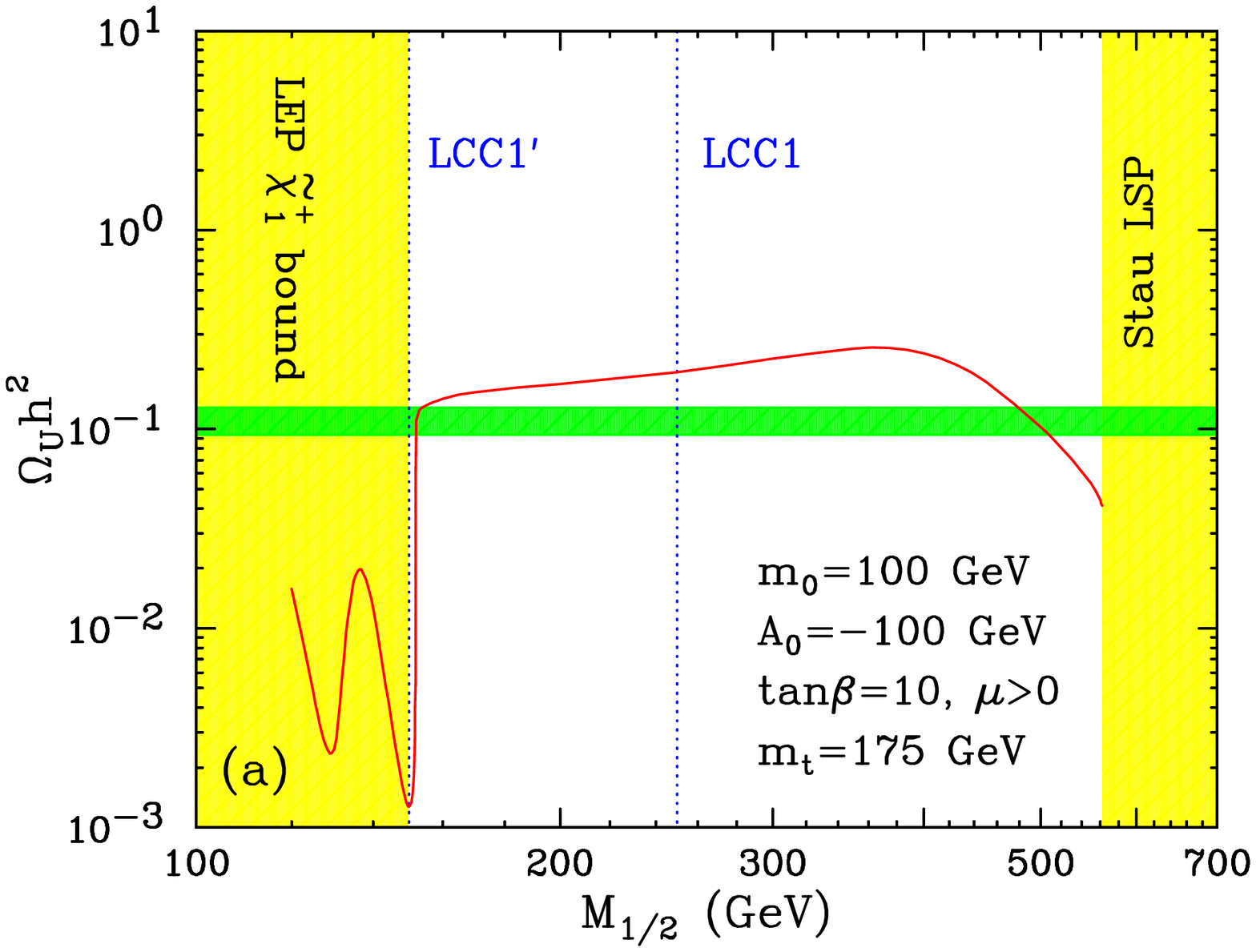,width=7.2cm}
\epsfig{file=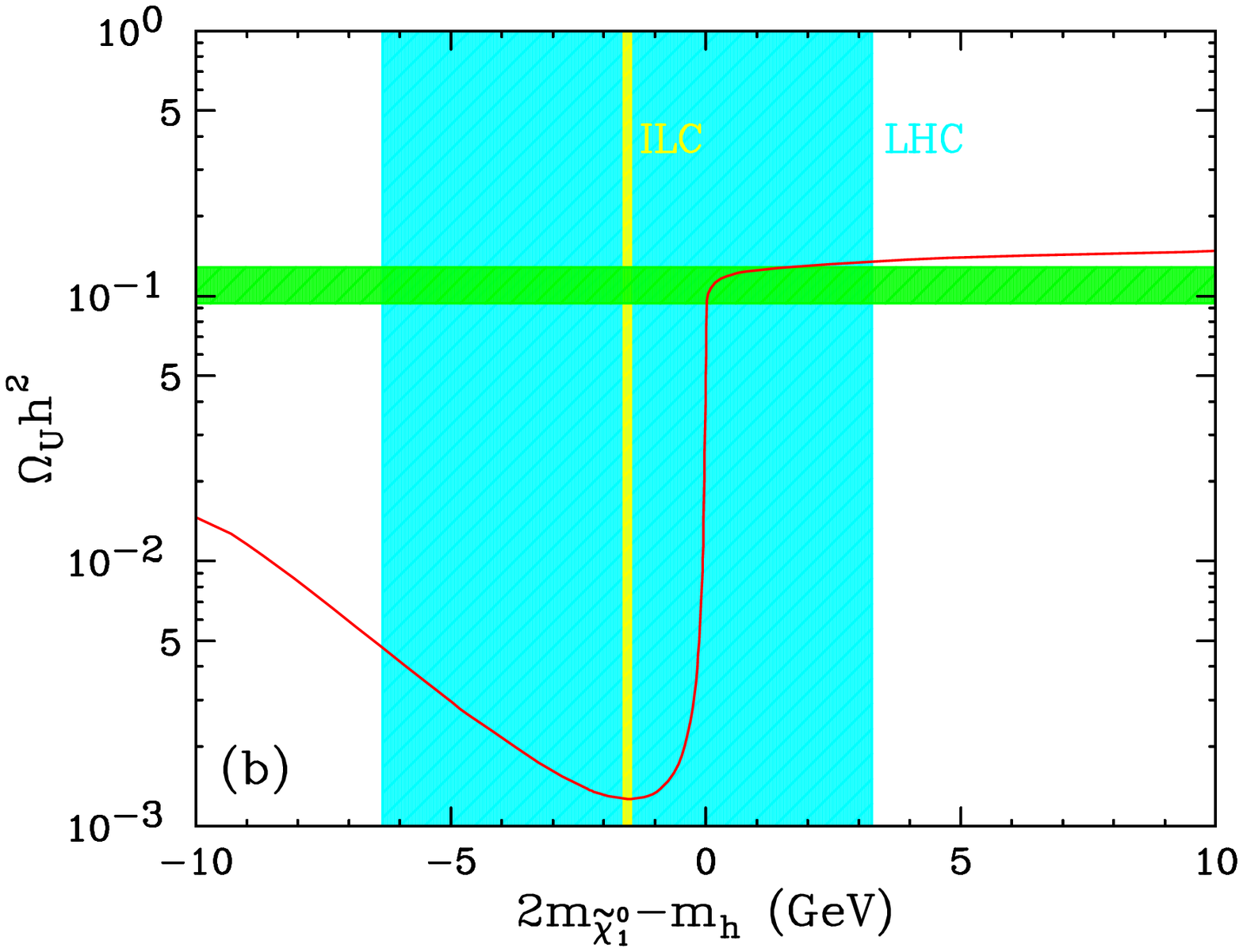,width=7.2cm}
\caption{\sl a) The dark matter relic abundance $\Omega_U h^2$
in standard cosmology, as a function of the universal 
gaugino mass parameter $M_{1/2}$ in minimal supergravity, for fixed
$m_0=100$ GeV, $A_0=-100$ GeV, $\tan\beta=10$, $\mu>0$ and $m_t=175$ GeV.
The horizontal (green) shaded band denotes the current 2$\sigma$ range for the 
experimental determination of the dark matter relic abundance.
The vertical (yellow) shaded band on the left is ruled out  by the negative
chargino searches at LEP. 
The other vertical (yellow) shaded band on the right is ruled out because the 
lightest superpartner (stau) is charged.
The vertical line marked LCC1 (LCC1$'$) denotes the $M_{1/2}$ value for the 
LCC1 (LCC1$'$) study point (see Table~\ref{tab:points}).
b) The same as a), but plotted versus the mass difference
$2m_{\tilde\chi^0_1}-m_h$, which indicates the proximity to the light Higgs pole.
The vertical light blue (yellow) band indicates the expected 
experimental precision in determining the value of the combination 
$2m_{\tilde\chi^0_1}-m_h$ at the LHC (ILC).} \label{fig:bulk}}

Fig.~\ref{fig:bulk}a exhibits three regions where the relic abundance is 
reduced below the WMAP level. On the right-hand side, at large $M_{1/2}$ 
and near the stau LSP limit, this is due to stau coannihilations (see Sec.~\ref{sec:coann}). 
On the left-hand side, at low $M_{1/2}$, we see two dips in $\Omega_U h^2$ 
due to resonant neutralino annihilations through 
a $Z$ (the left dip) and the light CP-even Higgs boson $h$ (the right dip).
The $Z$ resonance is already inside the excluded region but the $h$ resonance is 
still allowed.  That is where we chose our modified LCC1$'$ point,
marked by the vertical line labelled LCC1$'$.

The SUSY spectrum of point LCC1$'$ is given in Table~\ref{tab:points}.
It is somewhat lighter than the spectrum at LCC1, but the hierarchy of states
is similar, which allows us to assume the same precision in the 
sparticle mass determinations as for point LCC1. The chargino mass 
is 97.5 GeV, but (as discussed above) is still allowed due to the 
presence of a light electron sneutrino, which degrades the LEP chargino bound. 
One might be worried that the light CP-even Higgs boson mass 
is also lighter - 108.7 GeV, which is below the LEP Higgs mass limit.
Indeed such a light Higgs boson is ruled out. However, the exact value 
of the Higgs boson mass is not essential for our analysis, since we
are only interested in the {\em precision} with which the masses can 
be determined, rather than their actual values. At the original point LCC1
the problem was avoided by considering a heavier spectrum, which resulted 
in an unacceptable value for the relic density. Both of these problems 
can be simply solved by relaxing some of the mSUGRA assumptions, for example,
scalar mass nonuniversality in the third generation \cite{Baer:2004xx}
can easily lift the Higgs mass, while gaugino non-universality can remove 
any remaining tension with the light chargino bound from LEP.

Given the SUSY spectrum of point LCC1$'$, we now turn our attention to
the corresponding prediction for the WIMP relic density at colliders.
The most important feature of our spectrum is that $2m_{\tilde\chi^0_1}\approx m_h$,
which allows the neutralinos to annihilate very efficiently on the Higgs 
resonance. Unlike point LCC1, we now find that the dominant 
neutralino annihilation channels are $b\bar{b}$ (90\%) and $\tau^+\tau^-$ (9\%),
as expected on the $h$ resonance. This implies that the Higgs resonant diagram 
by far overwhelms all other neutralino annihilation processes {\em combined}.
This can already be guessed from Fig.~\ref{fig:bulk}a, where the turn on of the
resonant $h$ diagram causes a sharp drop in the relic density of almost 2 orders 
of magnitude. The same effect is even more evident in Fig.~\ref{fig:bulk}b, where
we plot our results from Fig.~\ref{fig:bulk}a versus the mass difference
$2m_{\tilde\chi^0_1}-m_h$, which is an indicator of the proximity to the 
light Higgs pole. Figs.~\ref{fig:bulk}a and \ref{fig:bulk}b make it clear
that in order to predict the neutralino relic abundance, we only need to
concentrate on the precision with which the masses $m_{\tilde\chi^0_1}$
and $m_h$ can be determined at colliders. At the LHC, $m_h$ can 
be precisely measured in the diphoton channel, therefore the dominant uncertainty
is still from the determination of $m_{\tilde\chi^0_1}$, which we take to 
be 5\%, as for point LCC1. At the ILC, we take the uncertainty on both 
$m_{\tilde\chi^0_1}$ and $m_h$ to be 0.05 GeV \cite{Baltz:2006fm}.
The resulting uncertainty on the combination $2m_{\tilde\chi^0_1}-m_h$
at the LHC (ILC) is shown in Fig.~\ref{fig:bulk}b by
the light blue (yellow) band. The remaining mass spectrum measurements
can be assumed to be similar to those for point LCC1, but as far as the
relic density is concerned, they become largely irrelevant, their 
only significance being to show that we are sufficiently far away from 
other special regions, e.g. the stau coannihilation region.

Fig.~\ref{fig:bulk}b can be used to directly translate the uncertainty 
on $2m_{\tilde\chi^0_1}-m_h$ into the corresponding uncertainty on
$\Omega_U h^2$, as calculated in standard cosmology. Turning on the 
effect of kination dominance, we then find the result in Fig.~\ref{fig:summary}a,
which shows the expected precision in the simultaneous determination of the
dark matter relic abundance $\Omega_K h^2$ and quintessence parameter 
$\eta_\Phi$ at the LHC (blue band) and the ILC (red band).
Fig.~\ref{fig:summary}a illustrates our main point -- that in kination 
dominated quintessence scenarios, colliders constrain {\em a combination}
of the dark matter and dark energy properties. In the case of point LCC1$'$,
measurements at the LHC alone will still be consistent with a
WIMP hypothesis within standard cosmology. One would really need the precision
of the ILC in order to see that within standard cosmology, this particular 
WIMP is not enough to explain all of the dark matter. Alternatively, if the
WIMP does make all of the dark matter in the universe, the ILC, combined with
precision cosmology, provide a measurement of the quintessence parameter 
$\eta_\Phi$ at the percent level.

\subsection{LCC2$'$: a study point in the focus point region}

In this and the next two subsections, we shall repeat the analysis we have 
just done for the LCC1$'$ bulk point, for our other three study points.
Study points in the bulk region of mSUGRA are relatively ``collider-friendly'', 
in the sense that a relatively large number of particles can be produced and 
studied, and as a result, a multitude of measurements can be made. 
The remaining three study points are chosen in regions where typically a smaller
set of states is accessible at colliders. Nevertheless, one can still obtain
a similar precision on the relic density determination, since the limited
knowledge of the spectrum is partially offset by the fact 
that only a few channels dominate the annihilation rate, and often it is only
a specific feature of the spectrum which governs the annihilation rate  
and thus needs to be known precisely.

The LCC2 study point was chosen in the so called ``focus point'' region
of mSUGRA \cite{Feng:1999mn,Feng:1999zg}. The region is characterized by 
relatively heavy scalars, which alleviates the tension with a number of 
phenomenological constraints: e.g. flavor problem, CP-violation, light Higgs 
boson mass, proton decay etc. \cite{Feng:2000bp,Feng:2005hw,Carena:2006gb}.
The LSP in the focus point region is still predominantly Bino, but
has a non-negligible Higgsino component, which opens up new 
annihilation channels into gauge and/or Higgs bosons \cite{Feng:2000gh}. 
For example, at LCC2 the LSP is 68\% Bino, 28\% Higgsino, and 4\% Wino, 
and the dominant annihilation channels are $W^+W^-$ (78\%),
$ZZ$ (12\%) and $Zh$ (6\%). The prediction of the relic density 
in this region is perhaps most sensitive to the amount of gaugino-higgsino 
mixing: increasing (decreasing) the Higgsino component of the LSP 
decreases (increases) the relic abundance. This is illustrated in 
Fig.~\ref{fig:foc}a, where we show the dark matter relic abundance 
$\Omega_U h^2$ in standard cosmology as a function of the universal 
scalar mass parameter $m_0$, for fixed
$M_{1/2}=300$ GeV, $A_0=0$ GeV, $\tan\beta=10$, $\mu>0$ and $m_t=175$ GeV.
\FIGURE[t]{
\epsfig{file=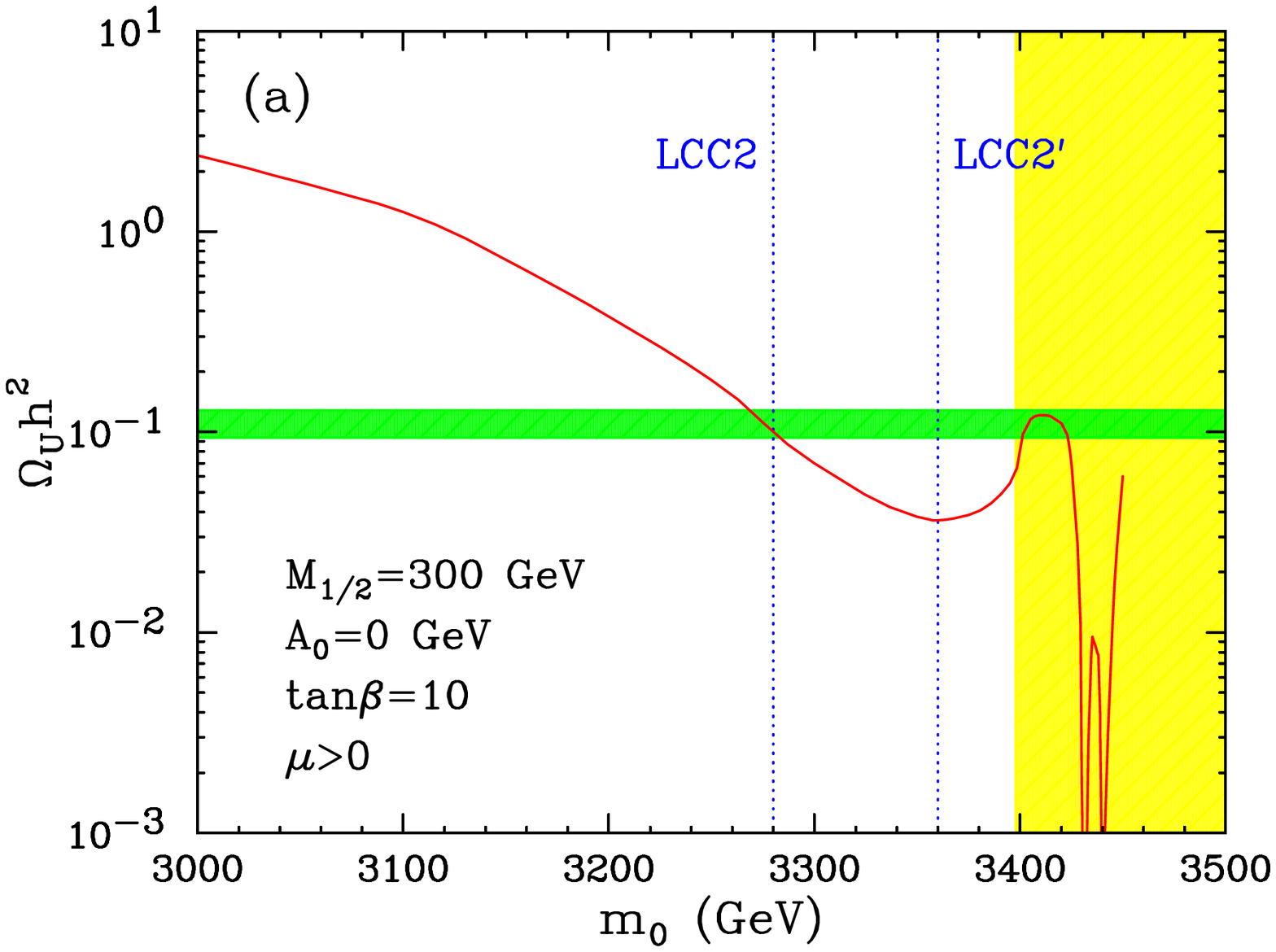,width=7.2cm}
\epsfig{file=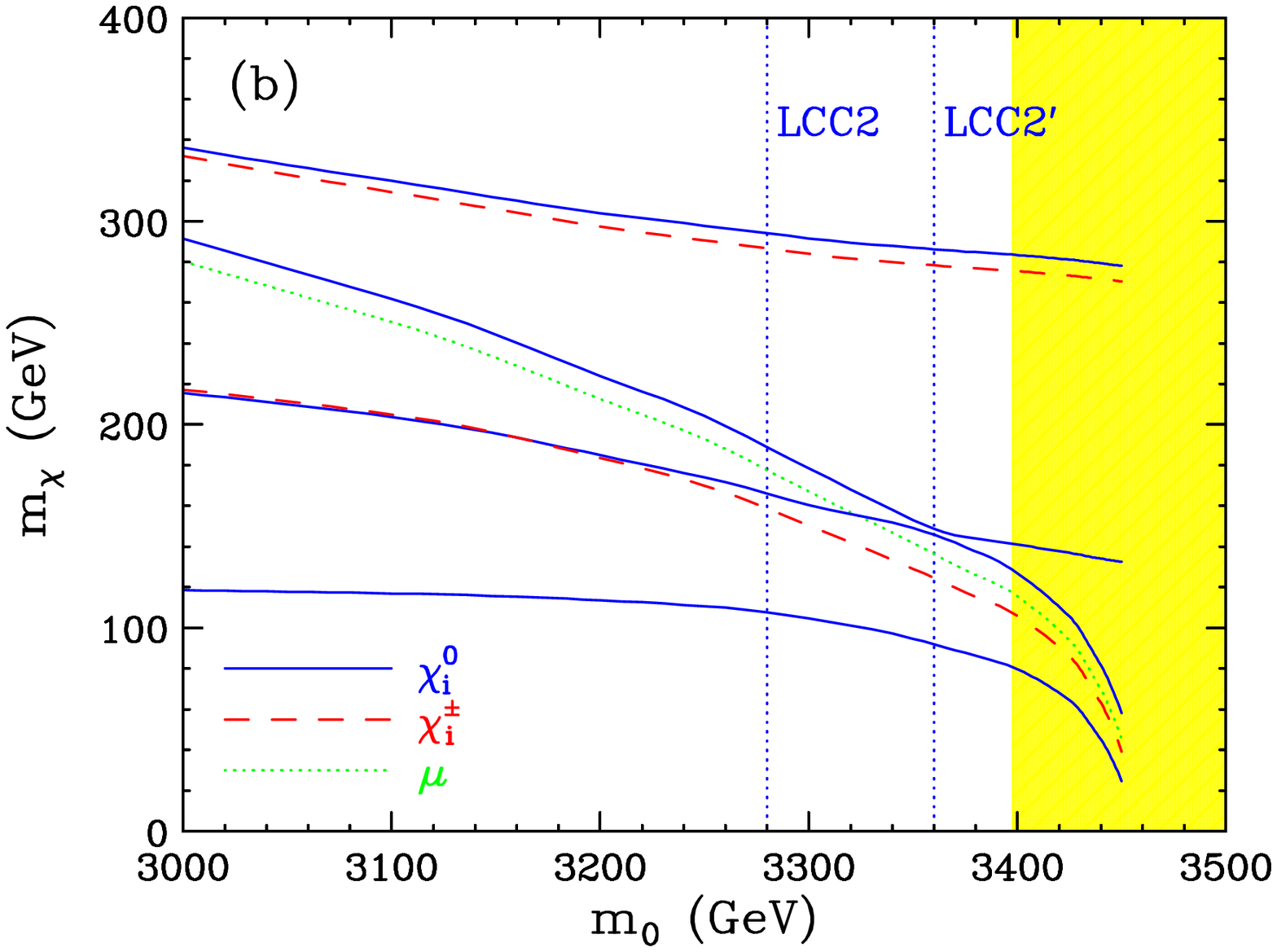,width=7.2cm}
\caption{\sl a) The dark matter relic abundance $\Omega_U h^2$
in standard cosmology, as a function of the universal 
scalar mass parameter $m_0$ in minimal supergravity, for fixed
$M_{1/2}=300$ GeV, $A_0=0$ GeV, $\tan\beta=10$, $\mu>0$ and $m_t=175$ GeV.
The horizontal (green) shaded band denotes the current 2$\sigma$ range for the 
experimental determination of the dark matter relic abundance.
The vertical (yellow) shaded band on the right is ruled out from the negative
chargino searches at LEP. 
The vertical line marked LCC2 (LCC2$'$) denotes the $m_0$ value for the 
LCC2 (LCC2$'$) study point (see Table~\ref{tab:points}).
b) The masses of the charginos (dashed, red) and neutralinos (solid, blue),
and the value of the $\mu$ parameter (dotted, green)  
as a function of $m_0$, for the same fixed parameters as in a).}
\label{fig:foc}}
As before, the horizontal (green) shaded band denotes the current 2$\sigma$ range for the 
experimental determination of the dark matter relic abundance.
The value of the Higgsino mass parameter $\mu$ is anti-correlated with $m_0$, 
as seen in Fig.~\ref{fig:foc}b, where we show the masses 
of the charginos (dashed, red) and neutralinos (solid, blue),
and the value of the $\mu$ parameter (dotted, green)  
as a function of $m_0$, for the same fixed parameters as in
Fig.~\ref{fig:foc}a. We see that increasing the value of $m_0$ 
away from the LCC2 point will lead to a smaller $\mu$, larger
Higgsino component in the LSP, and correspondingly a lower $\Omega_U h^2$.
This is the motivation behind our choice of the new value of $m_0=3360$ GeV 
for our modified study point LCC2$'$, as marked by the vertical line in 
Figs.~\ref{fig:foc}a and \ref{fig:foc}b. At point LCC2$'$, the LSP is well mixed:
$56\%$ Higgsino, $38\%$ Bino and $6\%$ Wino. The dominant annihilation channels are
still $W^+W^-$ (88\%) and $ZZ$ (5\%). If we increase $m_0$ even further, 
the LSP becomes purely Higgsino-like, and its mass begins to track 
the decreasing value of $\mu$, eventually closing the $W^+W^-$ and $ZZ$ annihilation channels,
which leads to the local increase in $\Omega_U h^2$ observed in Fig.~\ref{fig:foc}a.
While this increase is in principle sufficient to bring the prediction
for $\Omega_U h^2$ back to the desired level, it takes place inside a region
of parameter space (shaded in yellow in Fig.~\ref{fig:foc}) where the 
lightest (Higgsino-like) chargino is too light and is ruled out 
by LEP.\footnote{The chargino constraint can be evaded for larger values 
of $m_0$ and $M_{1/2}$, where the pure Higgsino LSP can become a viable 
dark matter candidate \cite{Chattopadhyay:2005mv}.}
Just as in Fig.~\ref{fig:bulk}a, inside the light chargino region 
we again find two sharp dips in $\Omega_U h^2$, which correspond to 
resonant annihilations on the $h$ and $Z$ pole, respectively.

Points LCC2 and LCC2$'$ have similar spectra. The sfermions are very heavy 
and will escape detection at the LHC and ILC. Gluino production at the LHC
will lead to long decay chains yielding jets, leptons and missing energy.
B-tagging can improve the sensitivity of the gluino search \cite{Mercadante:2005vx}
and allow for a gluino mass measurement at the LHC \cite{Baer:2007ya}.
However, in the focus point region, the relic density is determined primarily 
by the properties of the neutralino and chargino sectors, which will be 
mapped out relatively well at the ILC \cite{Moroi:2005nc,Gray:2005ci,Baer:2005ky}.
In particular, all but the heaviest chargino and neutralino states are accessible 
to the ILC at a center-of-mass energy of 500 GeV. Given the expected set of 
measurements available at point LCC2 \cite{Gray:2005ci}, 
\cite{Baltz:2006fm} estimates the resulting uncertainty in $\Omega_U h^2$ as 82\% at the LHC and
14\% at the ILC. Given that the SUSY spectrum at our point LCC2$'$ has only minor 
quantitative differences from the spectrum at LCC2, we expect a similar precision 
on $\Omega_U h^2$ at point LCC2$'$ as well. Using the same uncertainty levels
(82\% at LHC and 14\% at ILC), we show in Fig.~\ref{fig:summary}b
the resulting expectations for the simultaneous determination of the dark matter relic 
abundance $\Omega_K h^2$ and quintessence parameter $\eta_\Phi$ at point 
LCC2$'$ for the LHC (blue band) and the ILC (red band). Our result should be 
taken with a grain of salt, since we have not done a dedicated analysis of 
the spectroscopy measurements at point LCC2$'$. Nevertheless, it seems that 
already at the LHC one would be able to place a lower bound on the quintessence
parameter $\eta_\Phi$, and limit its value to within 2-3 orders of magnitude.
The ILC, in turn, will significantly narrow down the allowed range for $\eta_\Phi$
and within the context of our scenario, provide a unique probe of dark energy, 
unavailable by other means.

\subsection{LCC3$'$: a study point in the stau coannihilation region}
\label{sec:coann}

In most of the mSUGRA parameter space, the LSP is sufficiently lighter than 
the rest of the SUSY spectrum, so that the relic density is determined primarily 
by the LSP self-annihilation rate. However, mSUGRA also exhibits special regions 
where the LSP is sufficiently degenerate with another supersymmetric particle, so that 
both are present at freeze out and can affect the resulting dark matter relic density
both through their self-annihilations as well as their co-annihilations with each other.
In general, turning on coannihilations can lead to an increase or a decrease of the
relic density, depending on the type and properties of the coannihilating particle.
In mSUGRA the LSP is usually a Bino-like neutralino, whose self-annihilation rates 
are typically rather small, overclosing the universe. The presence of coannihilations 
would then typically tend to enhance the overall effective annihilation rate and 
lower the dark matter relic abundance. 

Point LCC3 was chosen in the region where the lightest neutralino is very close in mass
to the lightest tau slepton $\tilde\tau_1$ (see Table~\ref{tab:points}). Their
mass splitting was carefully adjusted so that the neutralino-stau coannihilations 
\cite{Ellis:1998kh}
would dilute the relic density precisely to the WMAP levels. In addition to the
usual neutralino annihilation channels 
$\tilde\chi^0_1\tilde\chi^0_1\to b\bar{b}$ (21\%),
$\tilde\chi^0_1\tilde\chi^0_1\to \tau^+\tau^-$ (12\%),
$\tilde\chi^0_1\tilde\chi^0_1\to \mu^+\mu^-$ and 
$\tilde\chi^0_1\tilde\chi^0_1\to e^+ e^-$ (5\% each);
we also have sizable coannihilation effects:
$\tilde\chi^0_1\tilde\tau_1\to h \tau$ (21\%),
$\tilde\chi^0_1\tilde\tau_1\to \gamma \tau$ (17\%)
and $\tilde\chi^0_1\tilde\tau_1\to Z \tau$ (6\%).
Naturally, reducing the neutralino-stau mass splitting even further would
enhance the coannihilation contribution, and drop $\Omega_U h^2$
below WMAP levels, which is what we need for our modified study point.
A simple way to control the $\tilde\chi^0_1$-$\tilde\tau_1$ mass splitting is 
provided by the parameter $m_0$ which affects the mass of $\tilde\tau_1$,
but not $\tilde\chi^0_1$. This is illustrated in Fig.~\ref{fig:stau}a 
where we show the dark matter relic abundance $\Omega_U h^2$
in standard cosmology, as a function of the universal 
scalar mass parameter $m_0$ in minimal supergravity, for fixed
$M_{1/2}=360$ GeV, $A_0=0$ GeV, $\tan\beta=40$, $\mu>0$ and $m_t=175$ GeV.
The horizontal (green) shaded band denotes the current 2$\sigma$ range for the 
experimental determination of the dark matter relic abundance.
\FIGURE[t]{
\epsfig{file=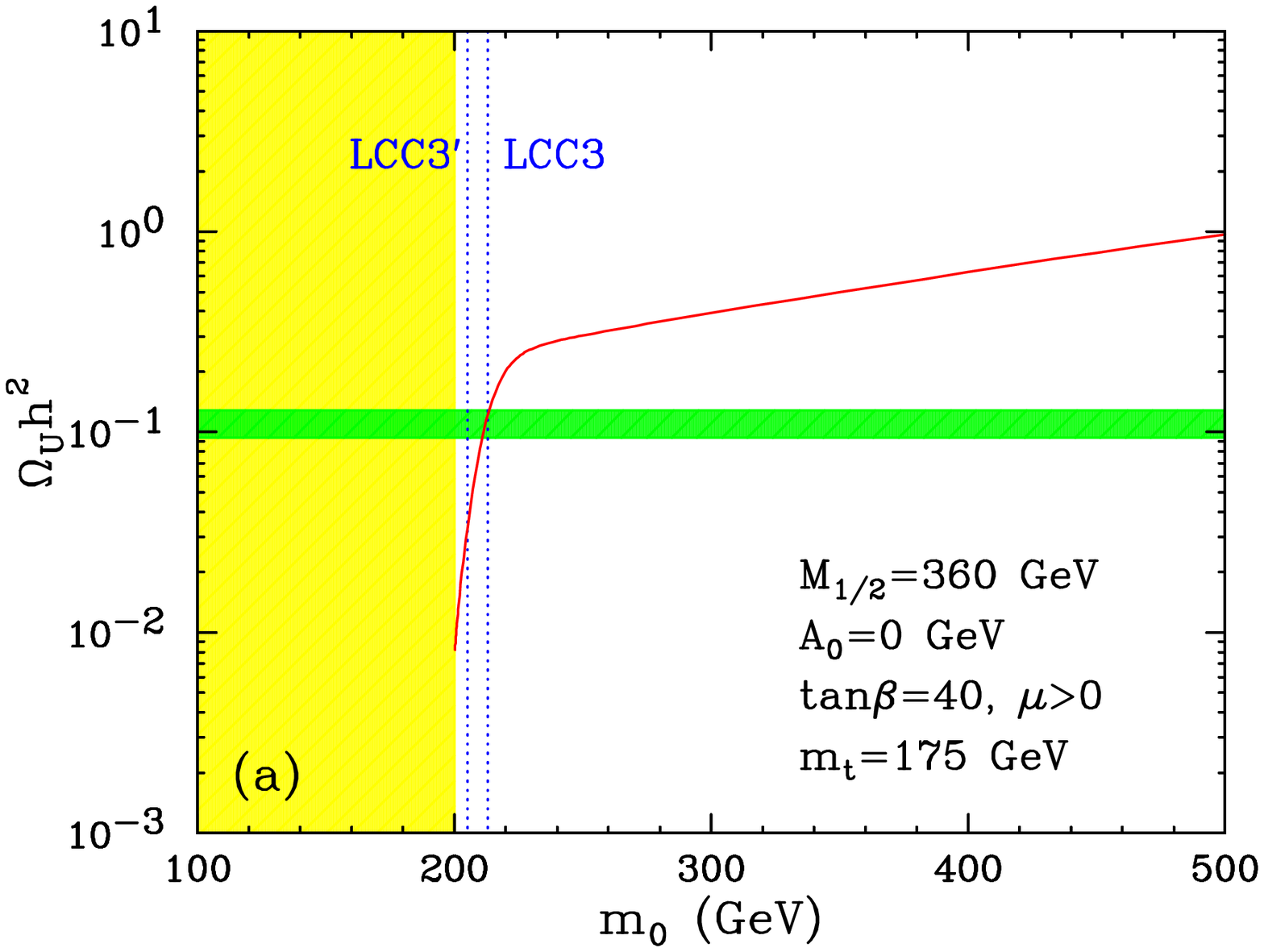,width=7.2cm}
\epsfig{file=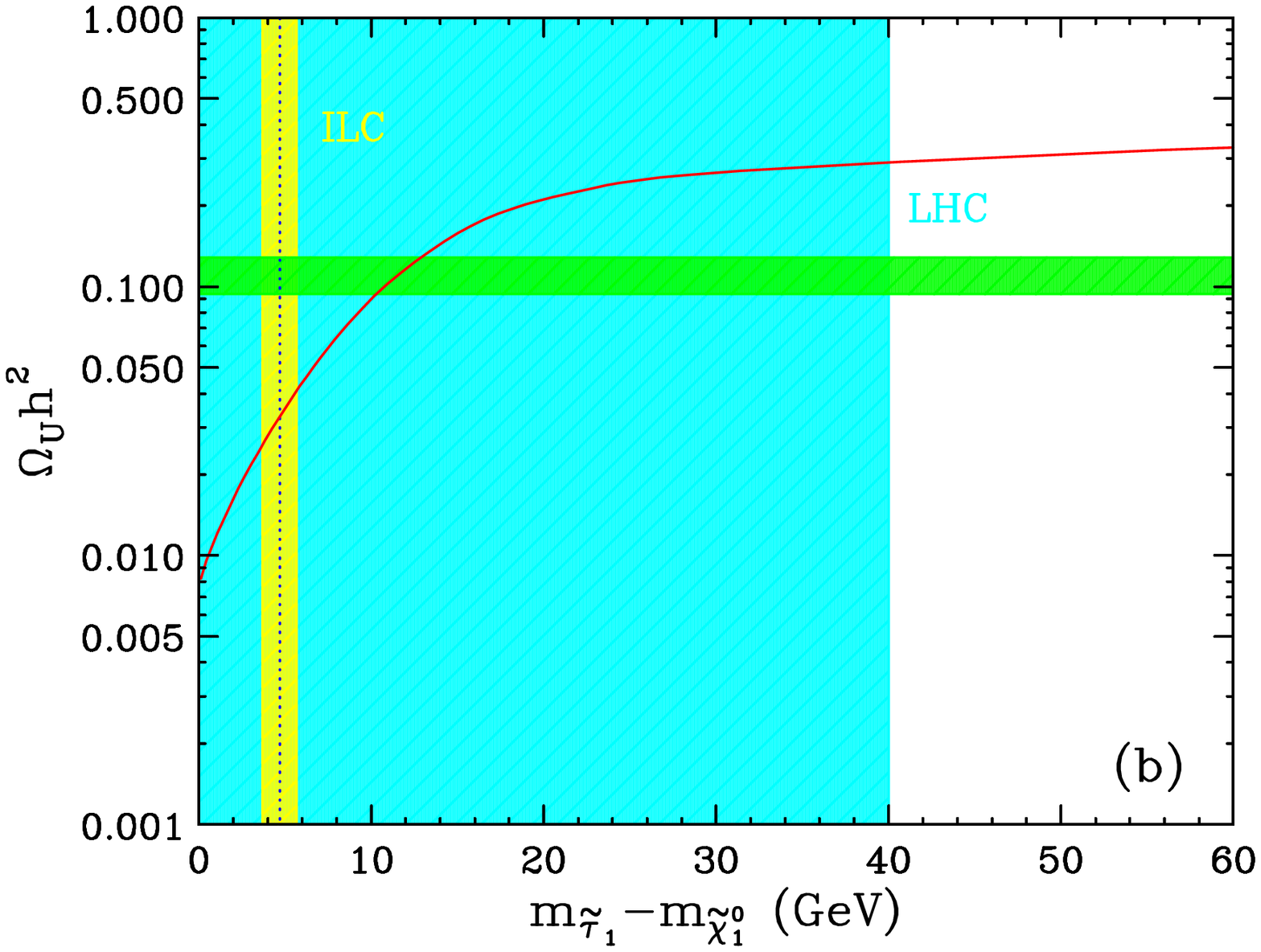,width=7.2cm}
\caption{\sl a) The dark matter relic abundance $\Omega_U h^2$
in standard cosmology, as a function of the universal 
scalar mass parameter $m_0$ in minimal supergravity, for fixed
$M_{1/2}=360$ GeV, $A_0=0$ GeV, $\tan\beta=40$, $\mu>0$ and $m_t=175$ GeV.
The horizontal (green) shaded band denotes the current 2$\sigma$ range for the 
experimental determination of the dark matter relic abundance.
The vertical (yellow) shaded band on the left is ruled out because the 
lightest superpartner (stau) is charged.
The vertical line marked LCC3 (LCC3$'$) denotes the $m_0$ value for the 
LCC3 (LCC3$'$) study point (see Table~\ref{tab:points}).
b) The same as a), but plotted versus the mass difference
$m_{\tilde\tau_1}-m_{\tilde\chi^0_1}$, which controls the effect of stau coannihilations.
The vertical light blue (yellow) band indicates the expected 
experimental precision in determining the value of the 
mass splitting
$m_{\tilde\tau_1}-m_{\tilde\chi^0_1}$ at the LHC (ILC).}
\label{fig:stau}}
The vertical line marked LCC3 (LCC3$'$) denotes the $m_0$ value for the 
LCC3 (LCC3$'$) study point (see Table~\ref{tab:points}).
The value of $m_0$ cannot be too low, since then 
$\tilde\tau_1$ becomes the LSP (in the vertical yellow-shaded band).
Nevertheless, close to the charged LSP region, the relic density 
gets to about an order of magnitude below the WMAP preferred value.
At point LCC3$'$, the neutralino-stau mass splitting is reduced from
12.3 GeV down to 4.7 GeV. This is sufficient to make coannihilations 
and stau annihilations dominate the relic density calculation.
At point LCC3$'$ we find that $\tilde\chi^0_1\tilde\chi^0_1$ annihilation 
processes only account for about 6\% of the total annihilation rate.
The dominant channels for dark matter number-changing processes are 
$\tilde\chi^0_1\tilde\tau_1\to h \tau$ (25\%),
$\tilde\chi^0_1\tilde\tau_1\to \gamma \tau$ (17\%),
$\tilde\tau^+_1\tilde\tau^-_1\to h h$ (17\%),
$\tilde\tau^+_1\tilde\tau^-_1\to \tau^+\tau^-$ (11\%), etc.

Precision spectroscopy at points LCC3 and LCC3$'$ is rather challenging at the LHC.
For one, the leptonic decay chain $\tilde\chi^0_2\to\ell^+\ell^-\tilde\chi^0_1$
is lost, as $\tilde\chi^0_2$ predominantly decays to taus. Furthermore, 
because of the small $\tilde\chi^0_1$-$\tilde\tau_1$ mass splitting, the taus 
from the $\tilde\tau_1$ decay tend to be relatively soft and 
difficult to reconstruct. A couple of recent analyses have attempted simultaneous
extraction of the neutralino-stau mass difference and the gluino mass
in a channel with jets, 2$\tau$'s and missing energy \cite{Arnowitt:2006jq} 
or a channel with jets, 3$\tau$'s plus missing energy \cite{Arnowitt:2006xw}.
However, it was found that those methods fail at very low neutralino-stau
mass differences -- below 5 GeV, as is the case of our point LCC3$'$. 
We shall therefore conservatively assume that in the absence of a definitive 
measurement, the LHC can only rule out large enough mass splittings ($\sim 40$ GeV)
which would have made a measurement possible. The ILC, on the other hand, can 
measure a neutralino-stau mass splitting as low as 5 GeV, down to about 
$\pm 1$ GeV \cite{Bambade:2004tq,Khotilovich:2005gb}. 
These uncertainties are used in Fig.~\ref{fig:stau}b, where
we plot the result from Fig.~\ref{fig:stau}a versus the mass difference
$m_{\tilde\tau_1}-m_{\tilde\chi^0_1}$, 
which controls the effect of stau coannihilations.
The vertical light blue (yellow) band indicates the expected 
experimental precision in determining the value of the 
mass splitting $m_{\tilde\tau_1}-m_{\tilde\chi^0_1}$ at the LHC (ILC),
as discussed above. We can now use the result from Fig.~\ref{fig:stau}b
to anticipate the precision (shown in Fig.~\ref{fig:summary}c) 
in the simultaneous determination of the
dark matter relic abundance $\Omega_K h^2$ and quintessence parameter 
$\eta_\Phi$ at point LCC3$'$, for the case of the LHC (blue band) and 
the ILC (red band). The challenges at the LHC mentioned earlier are readily 
evident, as the LHC results alone will still be consistent with standard cosmology, 
ruling out only the largest possible values of $\eta_\Phi$.
With the addition of the ILC, one is again able to pinpoint quite
accurately the quintessence parameter, and rule out the WIMP
scenario within standard cosmology.

\subsection{LCC4$'$: a study point in the Higgs funnel region}

Our final example is a study point illustrating the so called
``Higgs funnel'' case \cite{Baer:2000jj,Ellis:2001ms}. 
The SUSY spectrum at point LCC4 is somewhat similar
to the one at point LCC3, however, the relic density is controlled by
a different physics process. Here neutralino annihilations 
occur sufficiently close to the $A$ resonance, so that
the relevant parameters are the heavy Higgs masses and widths.
In this sense, the situation is similar to our modified 
``bulk'' point studied in Sec.~\ref{sec:bulk}, where neutralinos annihilated
near the light Higgs pole.

When neutralino annihilations occur exactly on the $A$ pole, 
the relic density is typically too low, since the 
annihilation cross-sections are suppressed only 
by the Higgs width instead of the Higgs mass. Conversely, when neutralinos 
annihilate far away from the resonant pole, the relic density is
determined by the other annihilation channels and is typically too large
(unless we are in one of the regions discussed in the previous three sections). 
Therefore, there are two special places, on {\em both} sides of the $A$ pole, 
where the relic density will be just right. The parameters for point LCC4 
were chosen so that neutralino annihilations take place in the right place 
below the $A$ resonance.
The dominant annihilation channels at point LCC4 are 
$\tilde\chi^0_1\tilde\chi^0_1\to b\bar{b}$ (78\%) and 
$\tilde\chi^0_1\tilde\chi^0_1\to \tau^+\tau^-$ (14\%),
indicative of the heavy Higgs branching fractions into fermion pairs. 

Following our usual procedure, we wish to modify the LCC4 point so that to
reduce the relic abundance. Since we are already in the vicinity 
of the $A$ pole, the simplest way to achieve this without leaving 
the funnel region, is to simply reduce the heavy Higgs masses so that
neutralino annihilations proceed on resonance. In mSUGRA, the Higgs 
masses are directly controlled by the $m_0$ parameter 
and therefore can be reduced by lowering $m_0$. 
Fig.~\ref{fig:funold} shows
the resulting variation of the dark matter relic abundance $\Omega_U h^2$ 
in standard cosmology, for fixed values of the remaining LCC4 parameters
($M_{1/2}=420$ GeV, $A_0=0$ GeV, $\tan\beta=53$, $\mu>0$ and $m_t=178$ GeV).
\FIGURE[t]{
\epsfig{file=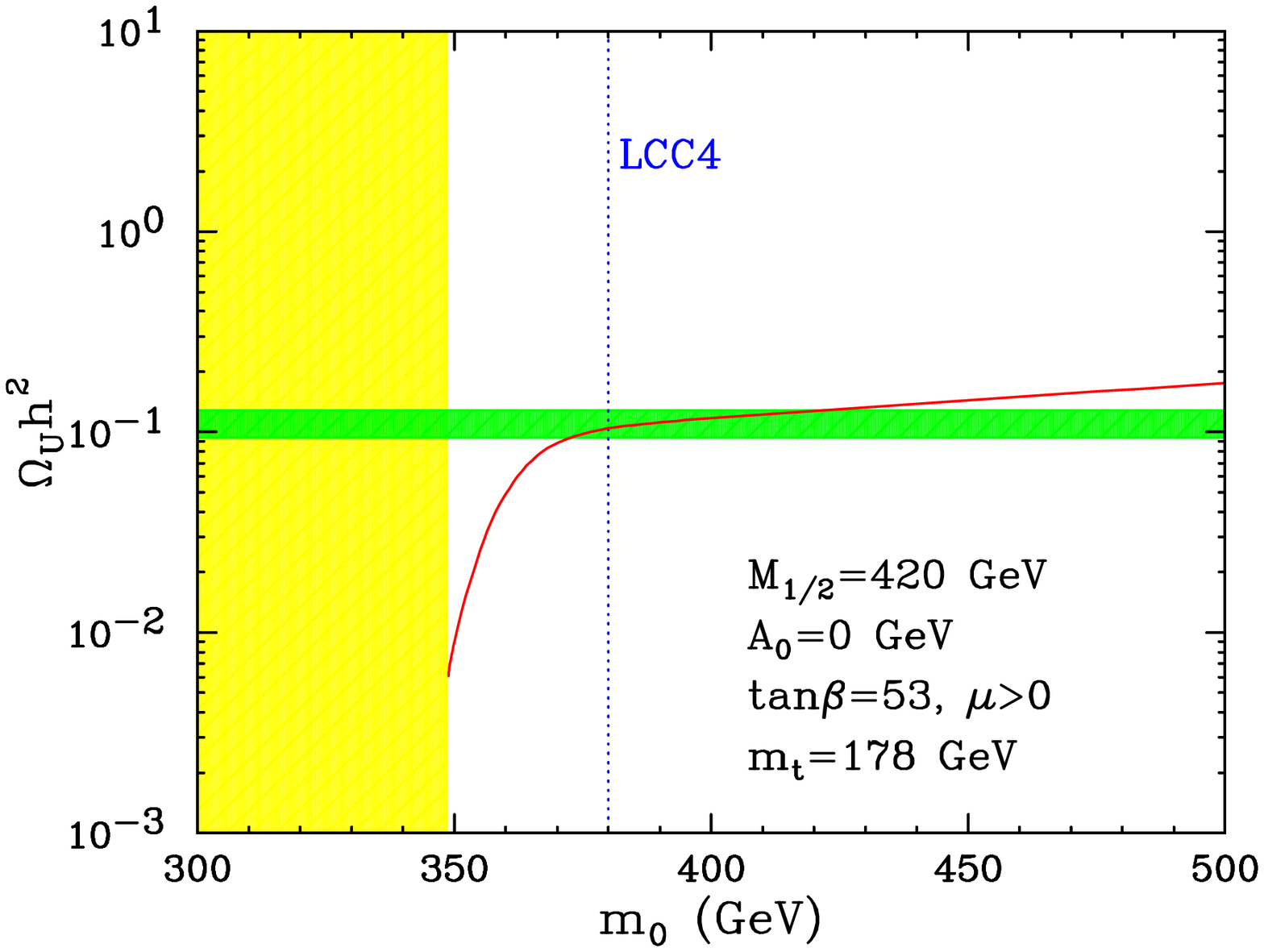,width=9.0cm}
\caption{\sl The dark matter relic abundance $\Omega_U h^2$
in standard cosmology, as a function of the universal 
scalar mass parameter $m_0$ in minimal supergravity, for fixed
$M_{1/2}=420$ GeV, $A_0=0$ GeV, $\tan\beta=53$, $\mu>0$ and $m_t=178$ GeV.
The horizontal (green) shaded band denotes the current 2$\sigma$ range for the 
experimental determination of the dark matter relic abundance.
The vertical (yellow) shaded band on the left is ruled out because the 
lightest superpartner (stau) is charged.
The vertical line marked LCC4 denotes the $m_0$ value for the 
LCC4 study point (see Table~\ref{tab:points}).
}
\label{fig:funold}}
As usual, the horizontal (green) shaded band denotes the current 
2$\sigma$ range for the 
experimental determination of the dark matter relic abundance.
The vertical line marked LCC4 denotes the $m_0$ value for the 
LCC4 study point (see Table~\ref{tab:points}). 
One might attribute the observed sharp reduction in the relic density 
at lower $m_0$ to approaching the heavy Higgs pole, however this is not 
really the case. In the vertical (yellow) shaded band on the left 
we find a stau LSP, hence this region is cosmologically ruled out.
Furthermore, near its boundary the neutralino and stau are quite 
degenerate, and we encounter the stau coannihilation situation discussed
in Sec.~\ref{sec:coann}. We have checked that the reduction in $\Omega_U h^2$
near the stau LSP boundary is primarily due to stau coannihilations as 
opposed to resonant annihilations on the Higgs pole (on the
stau LSP boundary we still find $m_A-2m_{\tilde\chi^0_1}\sim 75$ 
GeV). Therefore, the $m_0$
variation in the vicinity of the LCC4 study point does not reveal
the classic two-sided funnel shape, since the other half of the funnel 
is obscured by the stau coannihilation region. Similar conclusions 
hold if we vary $\tan\beta$ instead - just like $m_0$, $\tan\beta$ 
affects in a similar way both the Higgs and stau masses.
We therefore choose to select our point in a funnel region which is 
sufficiently far away from the stau coannihilation boundary.
This can be simply achieved by switching the sign of the $\mu$ parameter, 
which reverses the sign of the $\tan\beta$ enhanced Yukawa coupling 
corrections, leading to larger stau masses relative to the heavy Higgs masses.
We therefore modify the values of two parameters ($\tan\beta=50$ and $\mu<0$) 
and in Fig.~\ref{fig:fun}a once again we show the variation of the
dark matter relic abundance $\Omega_U h^2$ in standard cosmology, 
as a function of the universal scalar mass parameter $m_0$ in minimal supergravity.
We now clearly observe the classic Higgs funnel shape. The minimum of the
relic density is found right on resonance, around $m_0=940$ GeV, and 
the WMAP values (shown by the horizontal green shaded band) 
can be achieved on either side. The stau LSP region
is now sufficiently far away -- in fact the vertical yellow-shaded 
region on the left in Fig.~\ref{fig:fun}a is now ruled out because 
of a light Higgs boson in the spectrum. For our modified 
study point we choose the value $m_0=950$ GeV (denoted by the vertical blue line)
where we similarly find the dominant annihilation channels to be
$\tilde\chi^0_1\tilde\chi^0_1\to b\bar{b}$ (91\%) and 
$\tilde\chi^0_1\tilde\chi^0_1\to \tau^+\tau^-$ (9\%).

\FIGURE[t]{
\epsfig{file=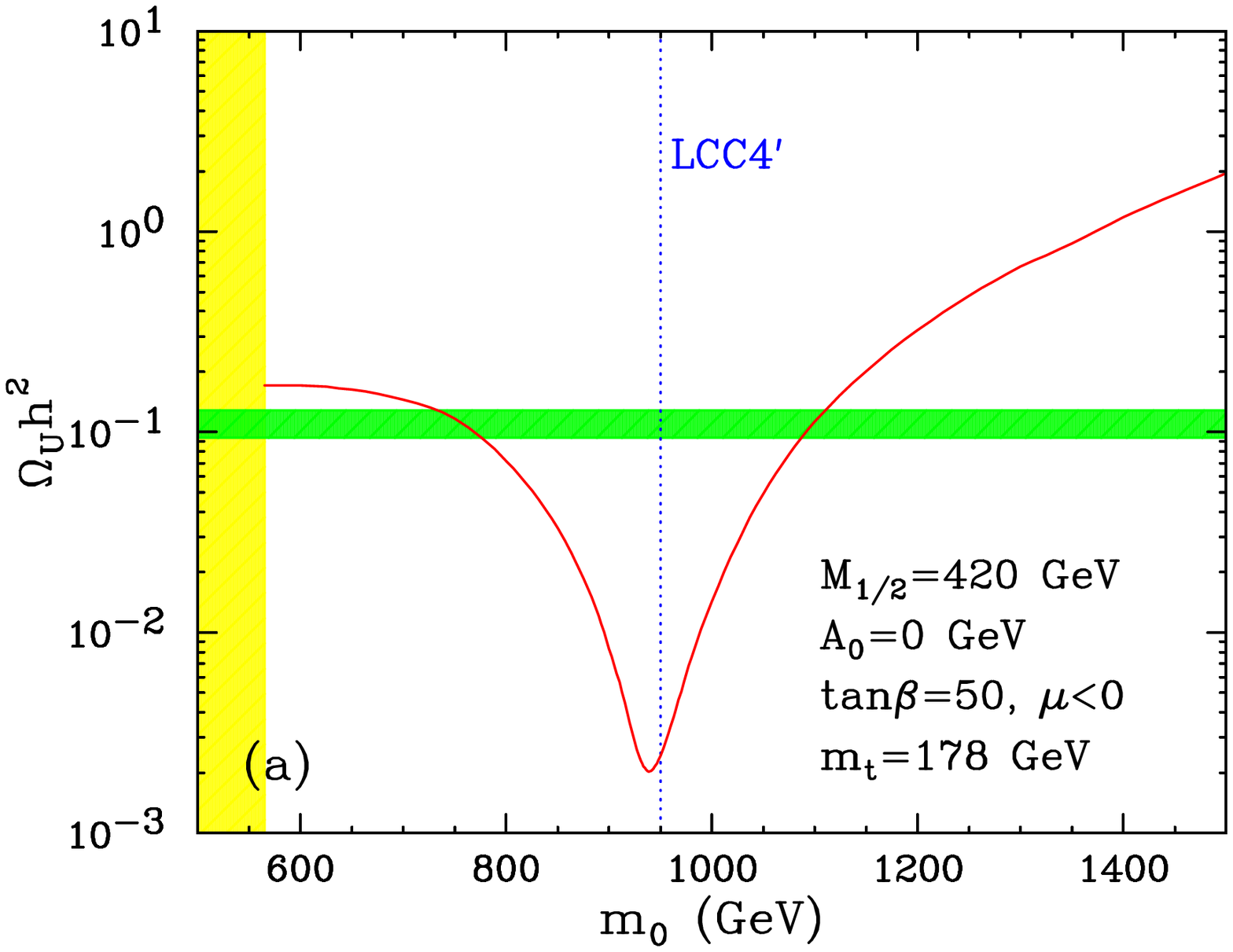,width=7.2cm}
\epsfig{file=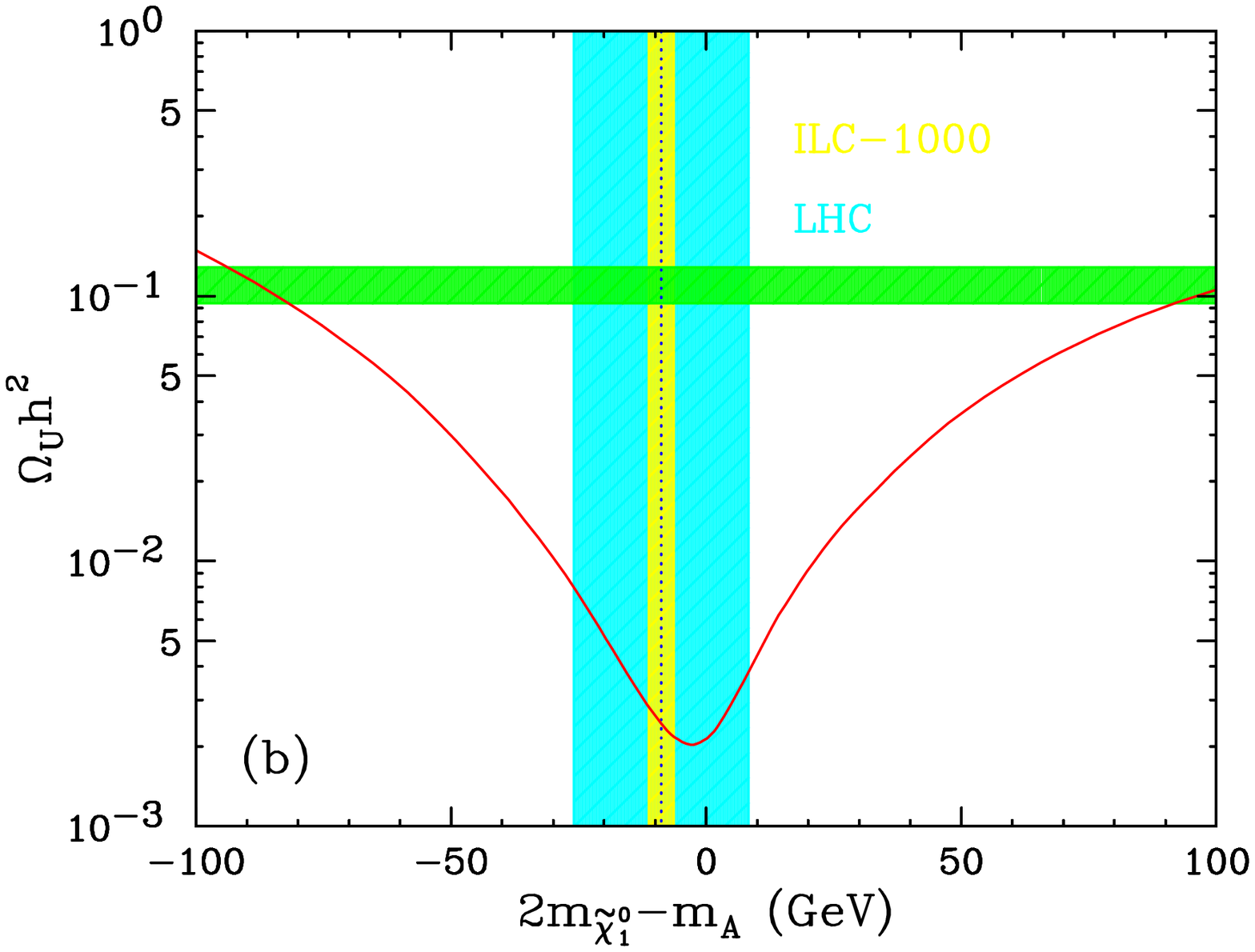,width=7.2cm}
\caption{\sl a) The dark matter relic abundance $\Omega_U h^2$
in standard cosmology, as a function of the universal 
scalar mass parameter $m_0$ in minimal supergravity, for fixed
$M_{1/2}=420$ GeV, $A_0=0$ GeV, $\tan\beta=50$, $\mu<0$ and $m_t=178$ GeV.
The horizontal (green) shaded band denotes the current 2$\sigma$ range for the 
experimental determination of the dark matter relic abundance.
The vertical (yellow) shaded band on the left is ruled out because there is
a very light Higgs boson in the spectrum.
The vertical line marked LCC4$'$ denotes the $m_0$ value for the 
LCC4$'$ study point (see Table~\ref{tab:points}).
b) The same as a), but plotted versus the mass combination
$2m_{\tilde\chi^0_1}-m_A$, which indicates the proximity to the heavy Higgs pole.
The vertical light blue (yellow) band indicates the expected 
experimental precision in determining the value of the combination 
$2m_{\tilde\chi^0_1}-m_A$ at the LHC (ILC-1000).}
\label{fig:fun}}

We are now in position to discuss our expectations for the 
precision of the dark matter and dark energy determinations 
at the LHC and ILC at point LCC4$'$. As with any such Higgs funnel point,
it is most important to measure the heavy Higgs spectrum in 
addition to the LSP mass. For consistency, we shall again use
the assumptions of \cite{Baltz:2006fm}, that at point LCC4 
the LHC will be able to determine the neutralino (heavy Higgs) mass 
to within 17 GeV (1.5 GeV). Clearly, the error on the relevant 
mass combination $2m_{\tilde\chi^0_1}-m_A$ is then dominated by the error 
on the LSP mass determination. Therefore, we can use the same error 
estimates for our modified point LCC4$'$, which has a very similar value of 
the LSP mass. In Fig.~\ref{fig:fun}b we again plot the dark matter relic 
abundance, for the same fixed parameters as in Fig.~\ref{fig:fun}a,
but this time versus the relevant mass combination $2m_{\tilde\chi^0_1}-m_A$ 
(compare to Fig.~\ref{fig:bulk}b). The vertical light blue 
band indicates the expected experimental precision in determining 
the value of the combination $2m_{\tilde\chi^0_1}-m_A$ at the LHC.
Unfortunately, at both point LCC4 and LCC4$'$ the heavy Higgs spectrum is 
too heavy to be observed by the ILC at center-of-mass energy of 500 GeV (ILC-500).
In addition, at our modified point LCC4$'$ the staus are also heavy enough 
to escape detection at the ILC-500. Since ILC-500 does not add anything new to the
dark matter determinations, following \cite{Baltz:2006fm}
we choose to consider the ILC upgrade at center-of-mass energy of 
1000 GeV (ILC-1000). There the chargino-neutralino sector becomes accessible
and the LSP mass is expected to be measured to within 1.4 GeV. The 
vertical yellow band in Fig.~\ref{fig:fun}b then
indicates the corresponding experimental precision 
at ILC-1000 in determining the value of the combination 
$2m_{\tilde\chi^0_1}-m_A$ at our modified point LCC4$'$. 

\FIGURE[t]{
\epsfig{file=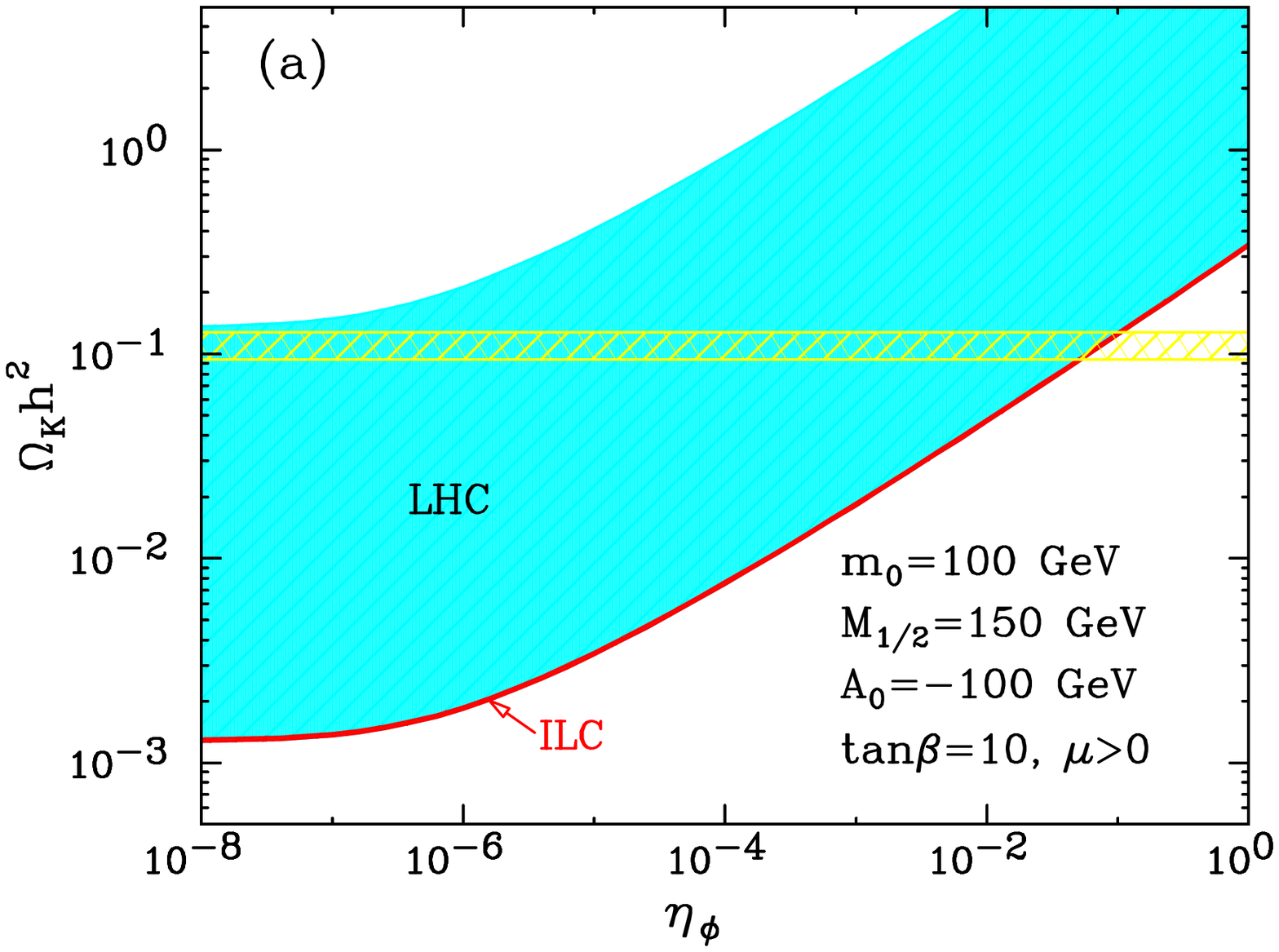,width=7.2cm}
\epsfig{file=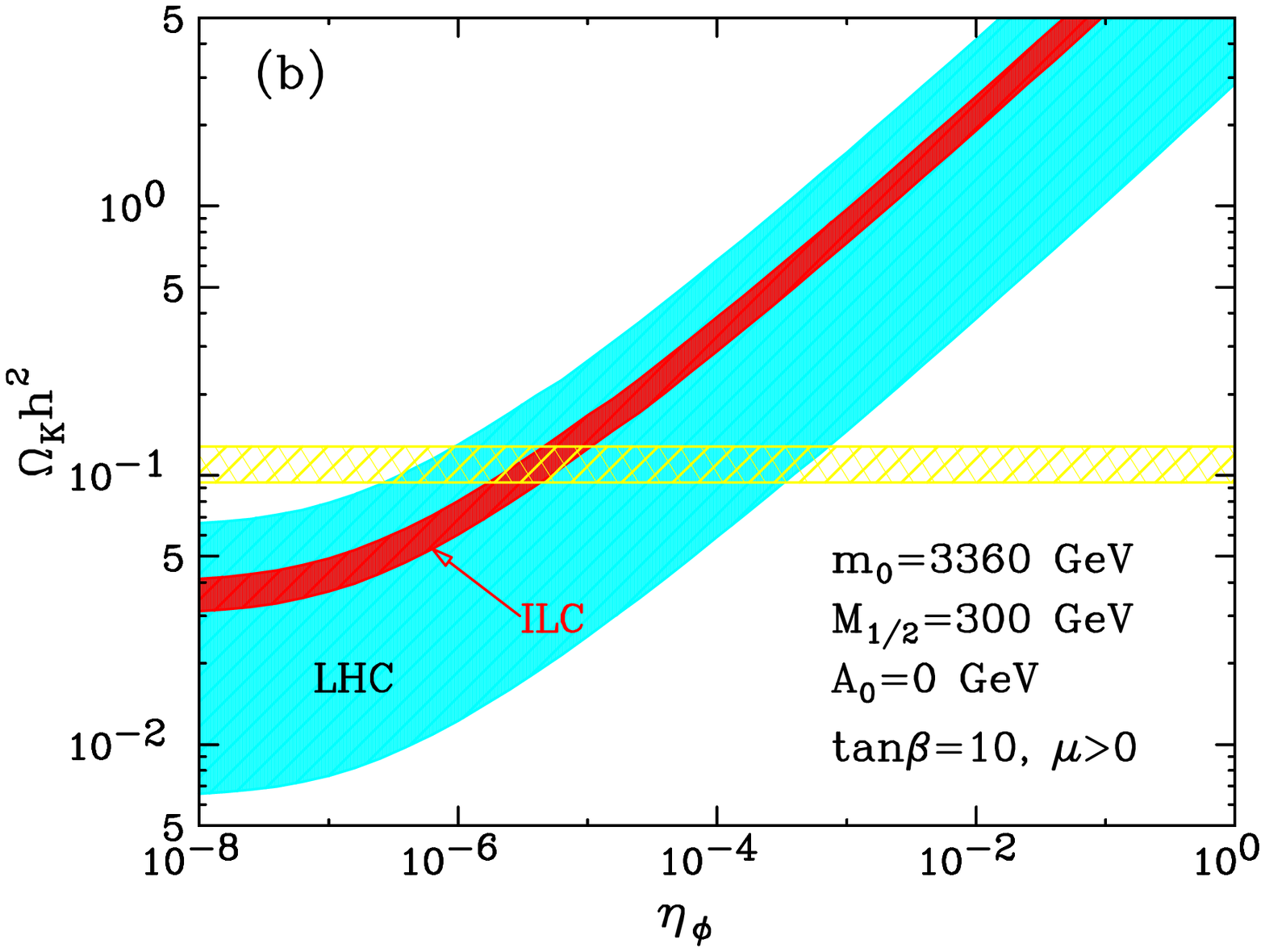,width=7.2cm}
\epsfig{file=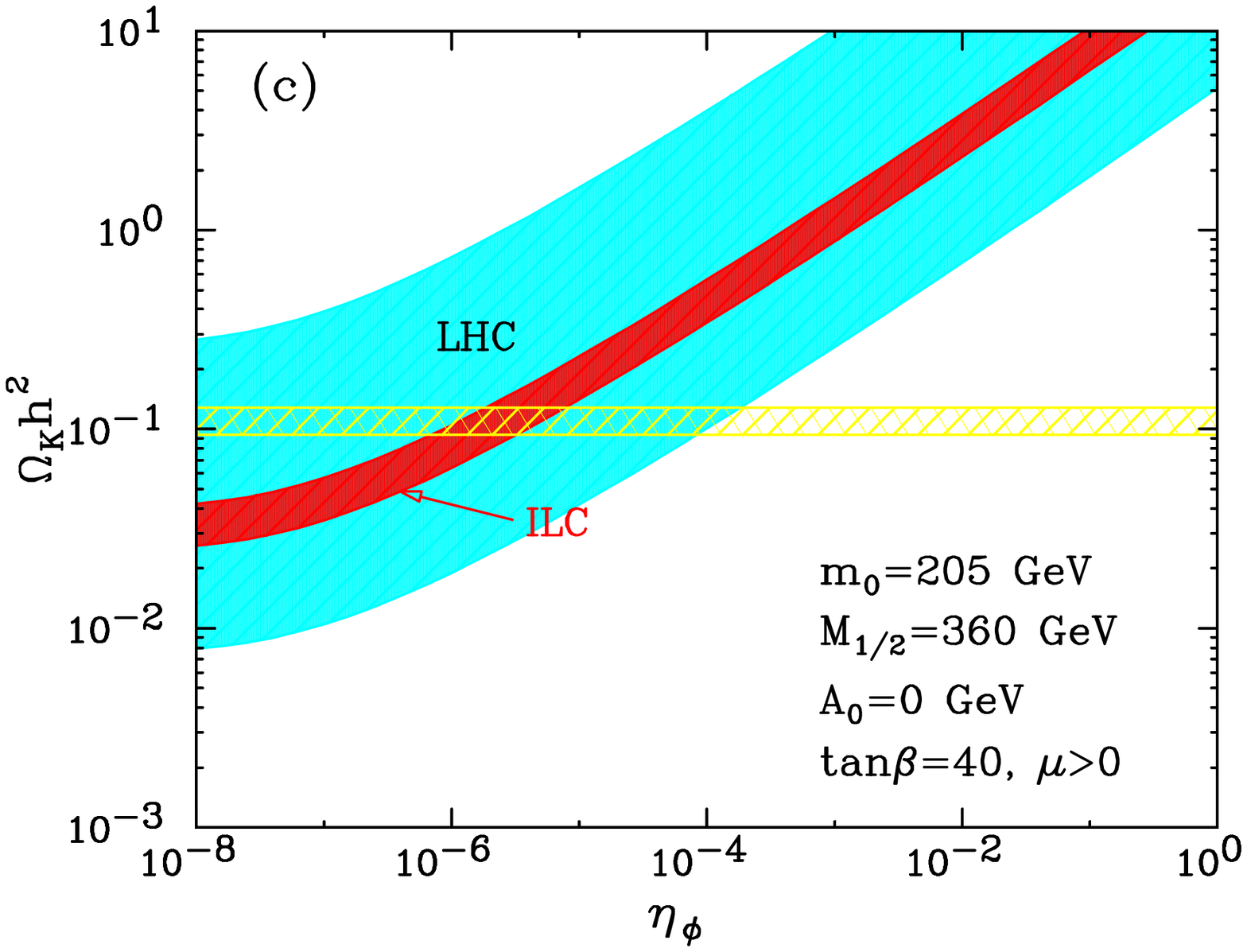,width=7.2cm}
\epsfig{file=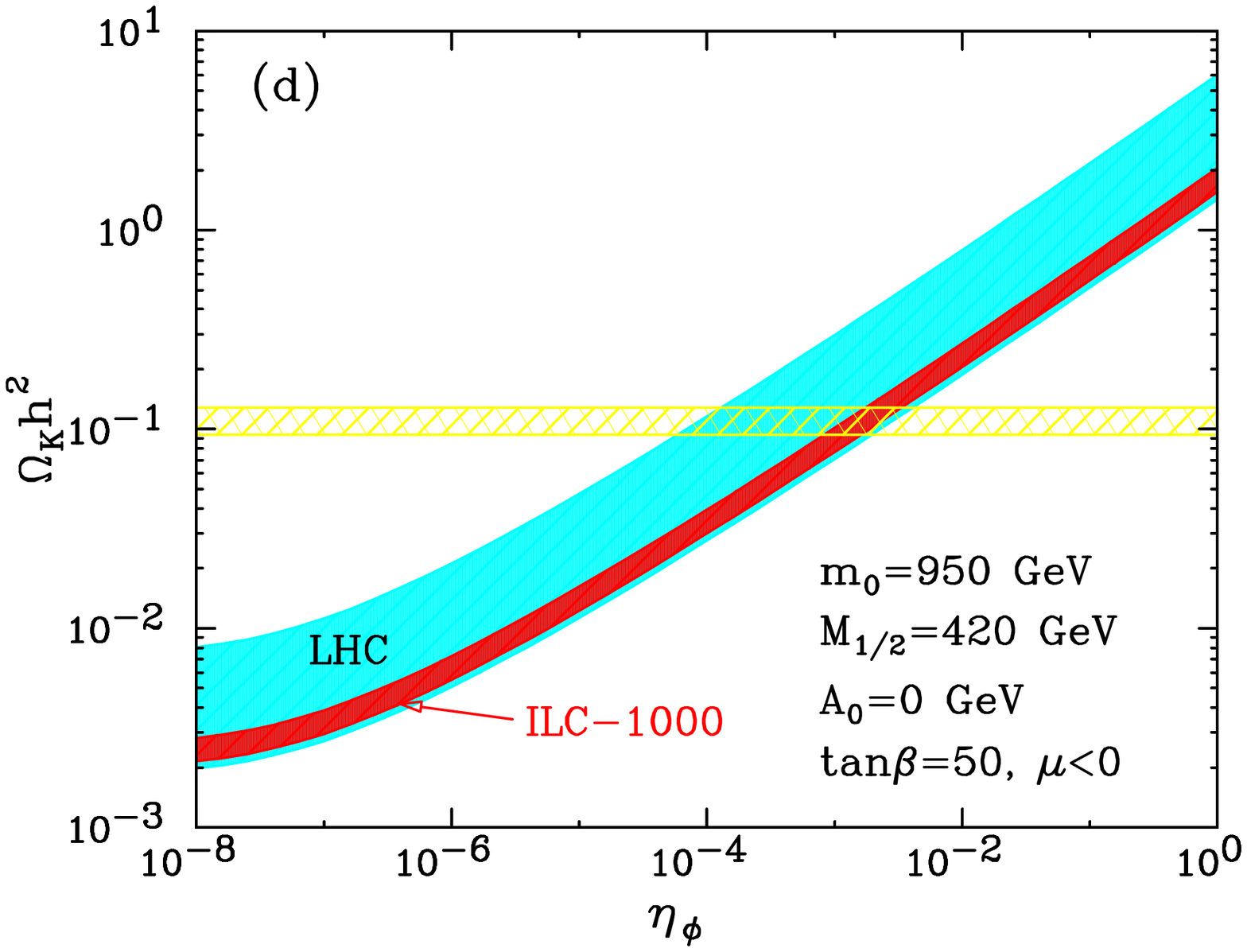,width=7.2cm}
\caption{\sl Expected precision in the simultaneous determination of the
dark matter relic abundance $\Omega_K h^2$ and quintessence parameter 
$\eta_\Phi$ at the LHC (blue band) and the ILC (red band), 
for a) LCC1$'$, b) LCC2$'$, c) LCC3$'$ and d) LCC4$'$ study point.}
\label{fig:summary}}

The experimental precision in determining the relevant sparticle properties 
from Fig.~\ref{fig:fun}b can now easily be translated into the corresponding
uncertainties on the dark matter and dark energy properties. The result 
for our modified point LCC4$'$ is shown in Fig.~\ref{fig:summary}d.
We see that, as in the case of LCC2$'$, the LHC can already 
uncover the mismatch between the measured WIMP particle properties
and those required by standard cosmology. The LHC can also provide a 
surprisingly good determination of the kination parameter $\eta_\Phi$ --
in fact, as can be seen by comparing the thickness of the LHC bands 
in the four panels of Fig.~\ref{fig:summary}, point LCC4$'$ is where 
the LHC does the best job. The advantage of the ILC as a precision 
machine is also readily seen in all 4 panels. For point LCC4$'$ 
where the ILC-500 energy is not sufficient, it should be
kept in mind that the ILC project will go ahead only after the first LHC 
results become available. Once the LHC observes the $A$ resonance 
near 350 GeV, and indicates that all electroweak and colored superpartners 
are rather heavy, it would be clear that the ILC design effort would
shift towards the higher energy ILC option. 

%%% DO NOT CHANGE ABOVE THIS LINE

\section{Discussion and Conclusions}
\label{sec:conclusions}

We are entering a new era in which the discovery of new fundamental
physics at colliders may help us answer some of the most important
puzzles of cosmology.  In particular, as the LHC is likely to lead to
the discovery of WIMPs that are identical to the particles which make
up dark matter galactic halos, we may soon learn about new
microphysics which governs the evolution of the universe.  Therefore,
it is natural to ask which aspects of cosmology can be probed by the
anticipated identification and microphysics of the dark matter particles.

One answer is that colliders will probe the period of our history at
temperatures of order of a few GeV when the dark matter particles
(assuming they are thermal WIMPs) presumably fell out of chemical thermal
equilibrium.  Therefore, we may be able to use collider information
to probe properties of every significant energy component in the
universe at the WIMP freeze out temperature, including field degrees
of freedom such as quintessence, which may be responsible for the dark
energy today.  Since
quintessence cannot be directly probed at any foreseeable terrestrial
controlled experiments, we have investigated to what extent LHC and ILC
can probe the cosmological properties of a very broad class of
quintessence models parameterized by a single parameter $\eta_\Phi
\lesssim 1$ (the ratio of energy density of quintessence to photon
energy density at the time of BBN defined as $T=1$ MeV), assuming that
there was a period of kination domination during the time of freeze
out.\footnote{A period of kination domination during which the scalar
field energy density dilutes as $a^{-6}$ is natural for $\eta_\Phi \sim
\mathcal O(1)$ because of BBN constraints.  The particular
parametrization assumes that kination domination continues until
temperatures of 1 MeV, but the results can be applied even to
models for which this does not occur by making a
model-dependent rescaling of the results.}

To compare the power of LHC and ILC in probing the early universe dark
energy cosmology, we have considered mSUGRA models with parameters
analogous to the set of four LCC study points \cite{Baltz:2006fm}
which were chosen to represent the four ``good'' dark matter regions.
We find that for the cases in which the dark matter annihilates
primarily through the lightest Higgs resonance and for the stau
coannihilation region, the LHC is only able to put an upper bound on the
parameter $\eta_\Phi$ characterizing the quintessence dynamics, while the ILC can put both an upper and a lower
bound on $\eta_\Phi$.  Since a lower bound means a
discovery of nontrivial dark energy dynamics while an upper bound is
consistent with no dark energy component, the ILC has the
potential to find supporting evidence for the possible kination
history of dark energy.  Indeed, it is remarkable that in the 
coannihilation region, the ILC can measure $\eta_\Phi$ as small as $10^{-6}$, while the 
LHC can only put an upper bound of $10^{-4}$ on the same quantity.  This should serve as yet one more important motivation for
the construction of the ILC.

Furthermore, such conjectures about the underlying dark energy dynamics
have other independent observable signatures which can be tested by
near-future non-collider experiments. For example, observable signatures
and correlated constraints were discussed in a previous publication
\cite{Chung:2007vz} for the broad class of kination-dominated
quintessence scenarios with an embedding in inflationary cosmology
relevant for our present study.  Therefore, just as big-bang
nucleosynthesis has served as a sturdy pillar of cosmology to constrain
dynamics of many conjectures of physics beyond the Standard Model, the
anticipated identification of dark matter at colliders and astrophysics
experiments will provide another robust pillar for cosmology, which will have 
profound consequences for uncovering the elusive nature of dark energy.

\begin{acknowledgments}
We would like to thank G. Kane for helpful conversations.  The work of 
DJHC was supported by DOE Outstanding Junior Investigator
Program through Grant No.  DE-FG02-95ER40896.  
The work of KTM was supported by DOE Outstanding Junior Investigator 
Program through Grant No.  DE-FG02-97ER41029. 
Fermilab is operated by Fermi Research Alliance, LLC under
Contract No. DE-AC02-07CH11359 with the Department of Energy.
\end{acknowledgments}

%%%%%%%%%%%%%%%%%%%%%%%%%%%%%%%%%%%%%%%%%%%%%%%%%%%%%%%%%%%%%%%
%\listoftables           % ONLY DRAFT
%\listoffigures          % ONLY DRAFT

%%%%%%%%%%%%%%%%%%%%%%%%%%%%%%%%%%%%%%%%%%%%%%%%%%%%%%%%%%%%%%%

\end{document}